\documentclass[lettersize,journal]{IEEEtran}
\usepackage{amsmath,amsfonts}
\usepackage{algorithmic}
\usepackage{algorithm}
\usepackage{array}
\usepackage[caption=false,font=normalsize,labelfont=sf,textfont=sf]{subfig}
\usepackage{textcomp}
\usepackage{stfloats}
\usepackage{url}
\usepackage{verbatim}
\usepackage{graphicx}
\usepackage{cite}
\usepackage{float}
\usepackage{bm}
\usepackage{amssymb}
\usepackage{color}
\usepackage{subfiles}

\newcommand{\nocolor}[1]{{\color{black}#1}}

\begin{document}

\title{Faster-Than-Nyquist Symbol-Level Precoding for Wideband Integrated Sensing and Communications}

\author{
  \IEEEauthorblockN{Zihan Liao},
  \and
  \IEEEauthorblockN{Fan Liu,~\IEEEmembership{Member,~IEEE}},
  \and
  \IEEEauthorblockN{Ang Li,~\IEEEmembership{Senior Member,~IEEE}},
  \and
  \IEEEauthorblockN{Christos Masouros,~\IEEEmembership{Senior Member,~IEEE}}
  \thanks{Z. Liao and F. Liu (corresponding author) are with the Department of Electrical and Electronic Engineering, Southern University of Science and Technology, Shenzhen, China. (email: liaozh2020@mail.sustech.edu.cn, liuf6@sustech.edu.cn).}
  \thanks{A. Li is with the School of Information and Communications Engineering, Faculty of Electronic and Information Engineering, Xi’an Jiaotong University, Xi’an, Shaanxi 710049, China. (email: ang.li.2020@xjtu.edu.cn).}
  \thanks{C. Masouros is with the Department of Electronic and Electrical Engineering, University College London, WC1E 7JE London, U.K. (e-mail: c.masouros@ucl.ac.uk).}
}



\maketitle

\begin{abstract}
In this paper, we present an innovative symbol-level precoding (SLP) approach for a wideband multi-user multi-input multi-output (MU-MIMO) downlink Integrated Sensing and Communications (ISAC) system employing faster-than-Nyquist (FTN) signaling. Our proposed technique minimizes the minimum mean squared error (MMSE) for the \nocolor{sensed} parameter estimation while ensuring \nocolor{the communication} per-user quality-of-service through the utilization of constructive interference (CI) methodologies. \nocolor{While the formulated problem is non-convex in general, we tackle this issue using proficient minorization and successive convex approximation (SCA) strategies.} Numerical results substantiate that our FTN-ISAC-SLP framework significantly enhances communication throughput while preserving satisfactory sensing performance.
\end{abstract}


\section{Introduction}
\IEEEPARstart{I}{ntegrated} sensing and communications (ISAC) has emerged as a pivotal enabling technology for next-generation wireless networks, such as 5G-Advanced (5G-A) and 6G. This technology seeks profound integration between wireless sensing and communication (S\&C) to facilitate the co-design of both functionalities, thereby enhancing hardware, spectral, and energy efficiency while obtaining mutual performance gains \cite{pin2021integrated}. As a result, \nocolor{ISAC has found applications in numerous emerging areas, including vehicular networks, industrial IoT, and smart homes \cite{zheng2019radar}.} 

Various \nocolor{signaling} schemes have been developed for ISAC, which can be broadly classified into two primary methodologies: \nocolor{orthogonal} resource allocation and fully unified waveform design. The former aims to allocate \nocolor{orthogonal} or orthogonal wireless resources to S\&C, thus to prevent interference between them, {\nocolor{namely, time-, spectral`-, spatial-, and code-division methods}}. However, this approach suffers from poor resource efficiency. Consequently, it is more advantageous to create a fully unified ISAC waveform through the shared utilization of wireless resources between S\&C. This strategy is generally referred to as Dual-Functional Radar-Communication (DFRC) design.

DFRC systems inherently exhibit conflicting requirements between radar and communication functionalities concerning aspects such as antenna placement, power amplifier operation regions, and signal formats, \nocolor{due to their different \nocolor{and often} contradictory performance metrics and constraints}. Thus, the transmit waveform must be meticulously designed to balance these requirements and enhance system performance. \nocolor{In general,} DFRC designs can follow one of three schemes: Sensing-centric design (SCD), communication-centric design (CCD), and joint design (JD) \cite{liu2022integrated}. The former two schemes prioritize the sensing or communication capabilities of the ISAC system, considering the other functionality as ancillary. In contrast, JD schemes strive to create an ISAC signal from scratch instead of relying on pre-existing S\&C waveforms, resulting in a scalable tradeoff between S\&C \cite{feng2020china}.

\nocolor{Recently,} multi-input multi-output (MIMO) architectures are extensively employed in DFRC systems to offer waveform diversity for radar target detection \cite{li2007mimo} and beamforming gains and spatial multiplexing for multi-user communications. Numerous researchers have focused on transmit precoding designs in MIMO DFRC systems \cite{mccormick2017simultaneous,qian2018joint,tang2020waveform,kumari2019adaptive,liu2020range,liu2018toward,liu2018mu,cheng2020hybrid,yuan2020bayesian,xu2020multi,su2020secure,bazzi2023outage}, \nocolor{where a precoding matrix is conceived to optimize radar sensing and communication metrics.} Notable radar metrics include radar receiver's signal-to-interference-plus-noise ratio (SINR) \cite{qian2018joint}, beampattern mean squared error (MSE) \cite{tang2020waveform}, Cram\'er-Rao bound \cite{kumari2019adaptive}, and the similarity between the designed DFRC beamformer and the its reference radar-only counterpart \cite{liu2020range,liu2018toward,liu2018mu,cheng2020hybrid}. Prevalent communication metrics encompass achievable rate \cite{yuan2020bayesian}, \cite{xu2020multi}, communication user's SINR \cite{liu2020range}, \cite{liu2018mu}, \cite{su2020secure}, and multi-user interference (MUI) \cite{tang2020waveform}, \cite{liu2018toward}. The amalgamation of radar sensing and communication metrics furnishes a comprehensive criterion for \nocolor{designing and evaluating} DFRC systems.

Although existing DFRC schemes are ingeniously conceived through sophisticated approaches, they generally assume Nyquist pulse shaping implicitly. Despite the tradeoff between S\&C performance, our objective in this paper remains to attain a substantial communication data rate without excessively compromising sensing performance. Consequently, faster-than-Nyquist (FTN) signaling is \nocolor{considered as a powerful technique} \cite{anderson2013faster}. FTN signaling's core concept is to enhance the data rate by accelerating transmitted pulses in the temporal dimension, thereby violating the Nyquist criterion and incurring controllable inter-symbol interference (ISI) \cite{anderson2013faster}. In a multi-user MIMO (MU-MIMO) system employing FTN signaling, interference manifests in both spatial and temporal domains, specifically, MUI and ISI due to non-orthogonality in users' channels and pulse shaping, \nocolor{which needs to be carefully coped with through multi-antenna precoding designs.}


In the aforementioned DFRC waveform designs \cite{mccormick2017simultaneous,qian2018joint,tang2020waveform,kumari2019adaptive,liu2020range,liu2018toward,liu2018mu,cheng2020hybrid,yuan2020bayesian,xu2020multi,su2020secure,bazzi2023outage}, conventional linear block-level precoding (BLP) embeds communication symbols into the dual-function waveform. However, these approaches' available degrees of freedom (DoFs) are proven to be limited by the number of users \cite{liu2020joint1,chen2020composite,liu2020joint2}. More \nocolor{importantly}, these methods \nocolor{adopted block-level beampattern, a function of the signal sample covariance matrix, as a design objective, where the radar sensing performance may be guaranteed only when the number of transmitted symbols is sufficiently large.} Consequently, instantaneous transmit beampatterns in different time slots might exhibit significant distortions, leading to severe performance degradation in target detection and parameter estimation if only a limited number of samples are collected. Additionally, conventional BLP designs \nocolor{mitigate} MUI and ISI via channel equalization techniques, such as zero forcing, which overlook the fact that known interference can be \nocolor{harnessed} to enhance useful signal power \cite{spano2018faster}.

To tackle the issues above, symbol-level precoding (SLP) has been proposed as a means of exploiting, rather than merely eliminating, interference in multi-user communication systems \cite{masouros2009dynamic,masouros2015exploiting,alodeh2018symbol,li2020tutorial,liu2020joint3,liu2021intelligent}. \nocolor{Particularly well-suited for ISAC applications, FTN signaling and SLP form an ideal pairing, as both spatial and temporal interference can be harnessed to enhance communication performance without compromising sensing performance.} Unlike conventional BLP, SLP is a non-linear and symbol-dependent approach, optimizing each instantaneous transmitted vector based on specific symbols to be transmitted. \nocolor{From a radar perspective,} this method enables \nocolor{meticulous} design of the instantaneous transmit beampattern in each time slot in a symbol-by-symbol manner, providing more DoFs for the sensing functionality. From a communication perspective, SLP can exploit transmitted symbol information to convert interference into constructive components, thereby enhancing the quality-of-service (QoS) of multi-user communications. \nocolor{Given the flexibility of JD-based DFRC waveform design, SLP can fully exploit constructive interference (CI).} 


\nocolor{Previous research on ISAC has predominantly adopted a narrowband model, with complex wideband communication tasks typically addressed through Orthogonal Frequency Division Multiplexing (OFDM) \cite{garmatyuk2009wideband}. However, as our work seeks to enhance data rates in the time domain as opposed to the frequency domain, the applicability of OFDM becomes less evident. This challenge is compounded by the expansion of the baseband signal's bandwidth under FTN signaling due to reduced symbol duration, thereby necessitating the consideration of a wideband communication model in our paper. While certain basic sensing tasks, such as angle detection, do not demand a wideband radar system, an increasing range of sensing requirements, including the detection of fast-moving objects and high range resolution detection, can only be adequately addressed using wideband radar systems \cite{immoreev2002ultra}. Therefore, to ensure comprehensive applicability and cater to the evolving requirements of our proposed ISAC system, we have opted to incorporate a wideband signal model in our research.}

In this paper, we propose a novel DFRC precoding technique referred to as FTN-ISAC-SLP for a MIMO ISAC system, wherein a multi-antenna BS simultaneously serves multiple single-antenna communication users and detects target response matrices for radar sensing. This approach amalgamates the strategies discussed above, thus actualizing performance enhancement for S\&C from both temporal and spatial dimensions. The existing literature on ISAC predominantly focuses on narrowband conditions, rendering the devised systems inapplicable in certain scenarios. In this paper, we extend the discussion to encompass wideband conditions, {\nocolor{and develop ISAC signaling strategies for wideband systems.}} The primary contributions of this work are summarized as follows:
\begin{itemize}
\item \nocolor{We develop the system model for wideband DFRC transmission using FTN signaling, and formulate the FTN-SLP waveform design \nocolor{as} an optimization problem}. By incorporating FTN signaling and SLP in the ISAC system, we aim to harness the benefits of both techniques, facilitating the exploitation of both temporal and spatial interference.
This combination is particularly suitable for ISAC applications and results in performance augmentation for sensing and communication from both temporal and spatial dimensions.
\item To efficiently solve the non-convex waveform design problem, we devise a pair of algorithm frameworks that employ minorization or SCA methods, which transform the problem into two solvable second-quadratically constrained quadratic programming (QCQP) sub-problems. The two approaches exhibit different performance. The minorization approach demonstrates rapid convergence, while the SCA approach excels in minimizing the objective function.
\item We further propose a more computationally efficient method, termed binary penalty search (BPS), to solve the sub-problems in minorization and SCA methods. The BPS method converts the QCQP into sequential quadratic programming (QP) problems, which can be readily solved with significantly reduced computational overheads.
\item \nocolor{We provide extensive numerical examples to illustrate the superiority of the proposed wideband FTN-ISAC-SLP designs over its Nyquist BLP counterparts, which demonstrate considerable performance gains in both radar sensing and multi-user communications.}
\end{itemize}

The remainder of this paper is structured as follows. Section II introduces the system model, the performance metrics for multi-user communications, and radar sensing, as well as the problem formulation. The proposed minorization and SCA algorithms, in addition to the BPS method, are developed in Section III. Simulation results are presented in Section IV, and conclusions are provided in Section V. Lastly, some proofs of the propositions in the paper are appended in Section VI.

\textit{Notation:} Boldface lower-case and upper-case letters indicate column vectors and matrices, respectively. $(\cdot)^{\top}$ and $(\cdot)^H$ denote the transpose and the transpose-conjugate operations, respectively. $\mathbb{C}$ denotes the set of complex numbers. $|a|$ and $\Vert b\Vert$ are the absolute value of a real scalar a and the magnitude of a complex scalar b, respectively. $\Vert\cdot\Vert_F$ is the Frobenius norm of a matrix argument. $\Re\left\{\cdot\right\}$, $\Im\left\{\cdot\right\}$ and $(\cdot)^{*}$ denote the real part, imaginary part and the conjugate of a complex number or matrix. $*$ denotes the convolution. $\otimes$ and $\circ$ denotes the Kronecker product and the Hadamard product, respectively. $\mathbf{A}\succeq \mathbf{0}$ indicates that the matrix $\mathbf{A}$ is positive semi-definite. $\mathbf{I}_M$ and 
 $\mathbf{0}_{M,N}$ indicate $M\times M$ identity matrix and $M\times N$ matrix with all entries being $0$, respectively. $\text{Diag}(\mathbf{a})$ denotes the diagonal matrix with its diagonal entries being the entries of the vector $\mathbf{a}$ respectively. $\text{diag}(\mathbf{A})$ denotes the vector with its entries being the diagonal entries of the matrix $\mathbf{A}$ respectively. $\mathbb{E}[\mathbf{A}]$ denotes the expectation matrix of the random matrix $\mathbf{A}$. $\text{tr}(\cdot)$ indicates the the trace of a square matrix. $\text{vec}(\mathbf{A})$ indicates the vector obtained by column-wise stacking of the entries of matrix $\mathbf{A}$. Finally, $\mathbf{a}\thicksim\mathcal{CN}(\mathbf{m},\mathbf{R})$ means that $\mathbf{a}$ obeys a complex Gaussian distribution with mean $\mathbf{m}$ and covariance matrix $\mathbf{R}$.

\section{System Model}
We consider a wideband MIMO ISAC BS equipped with $N_t$ transmit antennas and $N_r$ receive antennas, which is serving $K$ downlink single-antenna users while detecting targets as a monostatic radar. Without loss of generality, we assume $K < N_t$. Before formulating the FTN-ISAC-SLP problem, we first elaborate on the system model and performance metrics of both radar sensing and communications.
\begin{figure}[h]
    \centering
    \includegraphics[scale=0.5]{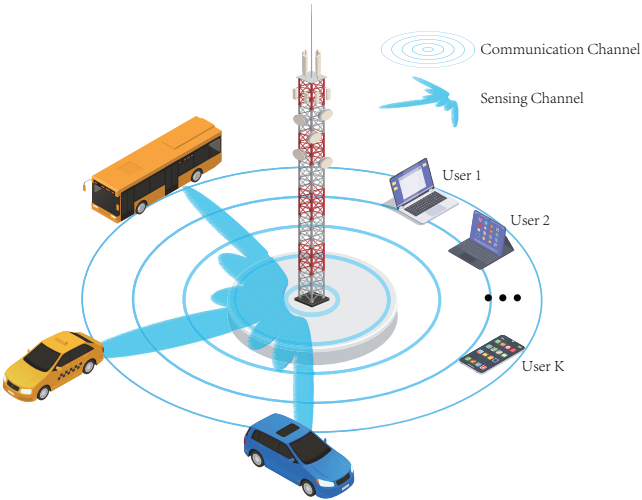}
    \caption{ISAC Downlink System.}
    \label{fig:system}
\end{figure}
\subsection{General Signal Model}
Let $\mathbf{D}=[\mathbf{d}_1, \mathbf{d}_2, \cdots, \mathbf{d}_{K}]^{\top} \in \mathbb{C}^{K \times L}$ denote the symbol matrix to be transmitted, with \nocolor{$\mathbf{d}_k \in \mathbb{C}^{L \times 1}$} being the symbol stream intended for the $k$-th user with a block length $L$, and each entry being drawn from a given constellation. Unless otherwise specified, in this paper we consider a PSK constellation, since the extension to QAM constellations is straightforward using approaches \nocolor{from the literature, for example those in} \cite{li2020tutorial}. Moreover, let $\mathbf{S} = [\mathbf{s}_1, \mathbf{s}_2, \cdots, \mathbf{s}_{N_t}]^{\top} \in \mathbb{C}^{N_t \times L}$ be the precoded signal matrix, with its entry $s_{n,i}$ at $n$-th row and $i$-th column representing the precoded symbols to be transmitted from $n$-th antenna at the $i$-th time slot. Suppose that the precoded symbols $\mathbf{S}$ are passed through a root-raised-cosine (RRC) shaping filter $\varphi(t)$ with a roll-off factor $\alpha$ and a duration $T_0$. The band-limited signal is transmitted with an FTN-specific symbol interval $T = \tau T_0$ where $\tau \in [0,1]$. Under such a setting, the transmitted FTN signal $x_n(t)$ at $n$-th antenna can be expressed as
\begin{equation}
    x_n(t) = \sum_{i=1}^{L}\varphi(t-(i-1)T)s_{n,i},
\end{equation}
where $s_{n,i}$ is $i$-th element of $\mathbf{s}_n$.

\subsubsection{Communication Model}
\nocolor{Suppose $h^{ij}(t)$ denotes the impulse response from the $j$-th transmitting antenna to the $i$-th receiving antenna. In the context of narrowband signaling, $h^{ij}(t)$ may be approximated as an impulse function, due to the frequency response of the channel being nearly constant within a narrow frequency range. However, in the case of wideband signaling, the same assumption becomes untenable, necessitating the consideration of a more general form for $h^{ij}(t)$. Consequently, the signal received at the $i$-th receiving antenna can be expressed as $y_i = \sum_j h^{ij}(t)*x(t) + n_i(t)$, where $*$ represents convolution. Given that $h^{ij}(t)$ can no longer be approximated as an impulse function, the convolution operation cannot be simplified to multiplication by a constant. Therefore, rather than directly using $\mathbf{y} = \mathbf{H}\mathbf{x} + \mathbf{n}$, we propose $\mathbf{y}(t) = \mathbf{H}(t) * \mathbf{x}(t) + \mathbf{n}(t) = \int \mathbf{H}(\tau)\mathbf{x}(t-\tau)d\tau + \mathbf{n}(t)$.}

Therefore the wideband MIMO input-output relationship in communication model is given by
\begin{equation}
    \mathbf{r}_c(t) = \mathbf{H}_c(t) * \mathbf{x}(t) + \mathbf{n}_c(t).
\end{equation}
where $\mathbf{n}_c(t)$ is the complex-valued AWGN with zero mean and variance $\sigma_C^2$, \nocolor{and the element} of the channel impulse response matrix $\mathbf{H}_c(t)$, \nocolor{namely, $h^{ij}_c(t)$}, is the impulse response from the $j$-th transmit antenna to the $i$-th receive antenna.

The received FTN signal after passing through a matched filter $\varphi^{*}(-t)$ is given by
\begin{equation}
\label{eq:c_model_1}
\begin{aligned}
    \mathbf{y}_c(t) &= (\varphi^*(-t)\mathbf{I}) * \mathbf{r}_c(t)  \\
    &= (\varphi^*(-t)\mathbf{I}) * \mathbf{H}_c(t) * \mathbf{x}(t) + (\varphi^*(-t)\mathbf{I}) * \mathbf{n}_c(t) \\
    &= \mathbf{H}_c(t) * (\phi(t)\mathbf{I}) * \mathbf{s}(t) + \bm{\eta}(t),
\end{aligned}
\end{equation}
where
\begin{equation}
\begin{aligned}
    \phi(t) &= \int_{-\infty}^{\infty}\varphi(\zeta)\varphi^{*}(\zeta-t)d\zeta, \\
    \bm{\eta}(t) &= \int_{-\infty}^{\infty}\mathbf{n}_c(\zeta)\varphi^{*}(\zeta-t)d\zeta.
\end{aligned}
\end{equation}
Let $\mathbf{X}_C=[\mathbf{x}_{C,1},\mathbf{x}_{C,2},\cdots,\mathbf{x}_{C,L}]$ and $\mathbf{x}_{C,i}$ be the sample of $(\phi(t)\mathbf{I}) * \mathbf{s}(t)$ at $t=(i-1)T$.

Then we can discretize $(\phi(t)\mathbf{I})*\mathbf{s}(t)$ with respect to time as 
\begin{equation}
\label{eq:discrete_xc}
    \mathbf{X}_C = \mathbf{S}\mathbf{\Omega}_{\phi}^{\top}.
\end{equation}
where 
\begin{equation}
\label{eq:define_omega_phi}
    \mathbf{\Omega}_{\phi} = 
    \begin{bmatrix}
        \phi(-QT) & 0 & \cdots & 0 \\
        \vdots & \phi(-QT) & \ddots & \vdots \\
        \phi(0) & \vdots & \ddots & 0 \\
        \vdots & \phi(0) & \ddots & \phi(-QT) \\
        \phi(QT) & \vdots & \ddots & \vdots \\
        \vdots & \phi(QT) & \ddots & \phi(0) \\
        \vdots & \vdots & \ddots & \vdots \\
        0 & 0 & \cdots & \phi(QT)
    \end{bmatrix}\in\mathbb{R}^{(L+2Q)\times L}.
\end{equation}

By letting $\mathbf{H}_C=[\mathbf{H}_{C,1},\mathbf{H}_{C,2},\cdots,\mathbf{H}_{C,P}]$ represent the sampled impulse response of $\mathbf{H}_c(t)$ at time $t=iT,i=0,1,\cdots,P-1$, $L_0=L+2Q$ and $L_1=L+2Q+P-1$, the equation (\ref{eq:c_model_1}) can be discretized as 
\begin{equation}
\label{eq:c_model_2}
    \widetilde{\mathbf{Y}}_C = \mathbf{H}_C*\mathbf{X}_C + \widetilde{\mathbf{N}}_C,
\end{equation}
where the $i$-th column of $\mathbf{H}_C*\mathbf{X}_C$ can be expressed as
\begin{equation}
\begin{aligned}
    &\mathbf{z}_{C,i} = \sum_j \mathbf{H}_{C,j}\mathbf{x}_{C,i-j}, \\
    &1\leq i\leq L_1,1\leq j\leq P,1\leq i-j\leq L_0.
\end{aligned}
\end{equation}
which is the result of matrix discrete convolution.

Notice that $\widetilde{\mathbf{Y}}_C\in\mathbb{C}^{K\times L_1}$. However, we need to recover $\mathbf{D}\in\mathbb{C}^{K\times L}$ from $\widetilde{\mathbf{Y}}_C$. To accomplish that, we right multiply a matrix $\mathbf{G}\in\mathbb{C}^{L_1\times L}$ to reduce the size of $\widetilde{\mathbf{Y}}_C$ to obtain the received symbol stream $\mathbf{Y}_C\in\mathbb{C}^{K\times L}$:
\begin{equation}
\label{eq:c_model_3}
    \widetilde{\mathbf{Y}}_C\mathbf{G} = (\mathbf{H}_C*\mathbf{X}_C)\mathbf{G} + \widetilde{\mathbf{N}}_C\mathbf{G}.
\end{equation}
One possible choice of $\mathbf{G}$ is $\mathbf{G}_1=[\mathbf{0}_{P+Q,L}^{\top},\mathbf{I}_L^{\top},\mathbf{0}_{Q-1,L}^{\top}]^{\top}$, which means we only take the first $L$ sample points and discard the rest ones. However, that would result in the incomplete use of the received signal energy. Another choice of $\mathbf{G}$ is 
\begin{equation}
    \mathbf{G}_2 =
    \begin{bmatrix}
        \mathbf{I}_{2L-L_1} & \mathbf{0}_{2L-L_1,L_1-L} \\
        \mathbf{0}_{L_1-L,2L-L_1} & \mathbf{I}_{L_1-L} \\
        \mathbf{0}_{L_1-L,2L-L_1} & \mathbf{I}_{L_1-L}
    \end{bmatrix},
\end{equation}
which means we add the last $L_1-L$ sample points to the previous points. However, that would result in the interference between \nocolor{consecutive symbols}. The design of $\mathbf{G}$ may be adjusted according to different scenarios and in this paper we take the first choice.

Moreover, we note that the noise $\widetilde{\mathbf{N}}_C\mathbf{G} = [\bm{\eta}_1, \bm{\eta}_2, \cdots, \bm{\eta}_K]^{\top}\mathbf{G}$, with $\bm{\eta}_k = [\eta_k(0), \eta_k(T), \cdots, \eta_k((L-1)T)]^\top$ being the corresponding received noise vector at the $k$-th user, is not independent at each time slot. 

\textit{Proposition 1:} For the noise $\bm{\eta}_k$ received at $k$-th user, we have
\begin{equation}
    \mathbb{E}[\bm{\eta}_k\bm{\eta}_k^H] = \sigma_C^2\mathbf{\Phi}_1,
\end{equation}
where 
\begin{equation}
\begin{aligned}
    &\mathbf{\Phi}_1 = \\
    &\begin{bmatrix}
        \phi(0) & \phi(-T) & \cdots & \phi(-(L_1-1)T) \\
        \phi(T) & \phi(0) & \cdots & \phi(-(L_1-2)T) \\
        \vdots & \vdots & \ddots & \vdots \\
        \phi((L_1-1)T) & \phi((L_1-2)T) & \cdots & \phi(0)
    \end{bmatrix}.
\end{aligned}
\end{equation}

\textit{Proof:} \nocolor{See section Appendix A}.

Thereby $\mathbb{E}[\mathbf{G}^{\top}\bm{\eta}_k\bm{\eta}_k^H\mathbf{G}] = \sigma_C^2\mathbf{G}^{\top}\mathbf{\Phi}_1\mathbf{G}$.
To decorrelate the noise, \nocolor{let the eigenvalue decomposition of $\mathbf{G}^{\top}\mathbf{\Phi}_1\mathbf{G}$ be} $\mathbf{U}_{\phi}\mathbf{\Lambda}_{\phi}\mathbf{U}_{\phi}^{H}$ where $\mathbf{U}_{\phi}$ is a unitary matrix containing eigenvectors and $\mathbf{\Lambda}_{\phi}$ is a diagonal matrix composed by eigenvalues. Right-multiplying $\mathbf{U}_{\phi}$ at both sides of (\ref{eq:c_model_3}) yields
\begin{equation}
\label{eq:c_model_4}
    \mathbf{Y}_C = (\mathbf{H}_C*\mathbf{X}_C)\mathbf{G}\mathbf{U}_{\phi} + \mathbf{N}_C,
\end{equation}
where $\mathbf{Y}_C=\widetilde{\mathbf{Y}}_C\mathbf{G}\mathbf{U}_{\phi}$ and $\mathbf{N}_C=\widetilde{\mathbf{N}}_C\mathbf{G}\mathbf{U}_{\phi}$.
By doing so, the covariance matrix for row vectors of $\mathbf{N}_C$ becomes $\sigma_C^2\mathbf{\Lambda}_{\phi}$, i.e., a diagonal matrix.

\subsubsection{Radar Sensing Model}
Consider the target response matrix (TRM) $\mathbf{H}_r(t) \in \mathbb{C}^{N_r \times N_t}$ that models the sensing channel. Depending on the sensing scenarios, $\mathbf{H}_r(t)$ can be of different forms.
According to the work of \cite{li2007mimo}, for angular extended target model where all the point-like scatterers are located in the same range bin, we have
\begin{equation}
    \mathbf{H}_r(t)=\sum_{i=1}^{N_s(t)}\alpha_i(t)\mathbf{b}(\theta_i(t))\mathbf{a}^H(\theta_i(t)),
\end{equation}
where $N_s(t)$ is the number of scatterers, $\alpha_i(t)$ and $\theta_i(t)$ denote the reflection coefficient and the angle of the $i$-th scatterer, and $\mathbf{a}\left(\theta\right) \in \mathbb{C}^{N_t \times 1}$ and $\mathbf{b}\left(\theta\right) \in \mathbb{C}^{N_r \times 1}$ are transmit and receive steering vectors.
Another example is to detect multiple point targets using OFDM waveforms \cite{sen2010multiobjective}. Then $\mathbf{H}_r(t)$ can be modeled as a TRM defined on the $n$-th subcarrier and the m-th OFDM symbol as
\begin{equation}
    \mathbf{H}_r(t)=\mathbf{B}(\Theta(t))\mathbf{C}_n(t)\mathbf{D}_m(t)\mathbf{A}^H(\Theta(t)),
\end{equation}
where $\mathbf{A}(\Theta(t))$ and $\mathbf{B}(\Theta(t))$ are transmit and receive steering matrices, and $\mathbf{C}_n(t)$ and $\mathbf{D}_m(t)$ are phase shifting matrices resulted by time delay and Doppler of each target.

\nocolor{In practical scenarios, however, the prior knowledge of targets, e.g., scatterer numbers and angle parameters of the to-be-sensed angular extended target may not be available at the BS}. As a result, there is typically no definitive structure for the sensing channel model. To ensure the generality of the proposed method, following \cite{liu2022integrated}, we consider a generic TRM $\mathbf{H}_R$ instead of a specific model. Subsequently, we may extract the parameters of the sensing channel from the estimated TRM. 

At the sensing receiver, we directly sample the received signal without passing it through the pulse-shaping filter, yielding the following radar received signal model. Similar to the equation (\ref{eq:c_model_1}), the received echo signal at the receive antennas can be written as
\begin{equation}
\label{eq:r_model}
    \mathbf{y}_r(t) = \mathbf{H}_r(t) * (\varphi(t)\mathbf{I}) * \mathbf{s}(t) + \mathbf{n}_r(t).
\end{equation}
where $\mathbf{n}_r(t)$ is the complex-valued AWGN at the receive antennas with zero mean and variance $\sigma_R^2$. We can then discretize $(\varphi(t)\mathbf{I}) * \mathbf{s}(t)$ in the same way with equation (\ref{eq:discrete_xc}) as
\begin{equation}
\label{eq:discrete_xr}
    \mathbf{X}_R = \mathbf{S}\mathbf{\Omega}_{\varphi}^{\top}.
\end{equation}
where 
\begin{equation}
\label{eq:define_omega_varphi}
    \mathbf{\Omega}_{\varphi} = 
    \begin{bmatrix}
        \varphi(-QT) & 0 & \cdots & 0 \\
        \vdots & \varphi(-QT) & \ddots & \vdots \\
        \varphi(0) & \vdots & \ddots & 0 \\
        \vdots & \varphi(0) & \ddots & \varphi(-QT) \\
        \varphi(QT) & \vdots & \ddots & \vdots \\
        \vdots & \varphi(QT) & \ddots & \varphi(0) \\
        \vdots & \vdots & \ddots & \vdots \\
        0 & 0 & \cdots & \varphi(QT)
    \end{bmatrix}.
\end{equation}
Then similar to equation (\ref{eq:c_model_2}) we can rewrite equation (\ref{eq:r_model}) as
\begin{equation}
\label{eq:r_model_1}
    \mathbf{Y}_R = \mathbf{H}_R*\mathbf{X}_R + \mathbf{N}_R,
\end{equation}
where $\mathbf{H}_R=[\mathbf{H}_{R,1},\mathbf{H}_{R,2},\cdots,\mathbf{H}_{R,P}]$ and $\mathbf{H}_{R,i}$ is the sampled TRM of $\mathbf{H}_r(t)$ at $t=(i-1)T$ and $\mathbf{N}_R$ denotes an AWGN matrix, with zero mean and the variance of each entry being $\sigma_R^2$. Here we assume every entry of $\mathbf{H}_R$ follows complex Gaussion distribution $\mathcal{CN}(0,\sigma_H^2)$.

\textit{Remark:} In the communication model we attempt to detect the signal $\mathbf{D}$ from $\mathbf{r}_c(t)$ in the receiver side, thus we pass the received signal to RRC matched filter to maximize the received SINR for each precoded symbol. In the sensing model our aim is to recover the TRM $\mathbf{H}_R$ from the raw observation (\ref{eq:r_model}), rather than to recover $\mathbf{D}$. Therefore, we treat $\mathbf{X}_R$ as an equivalent transmitted waveform and regard (\ref{eq:r_model}) as the sufficient statistics for estimating $\mathbf{H}_R$, which needs not to be match-filtered by the RRC pulse.

\subsection{Constraints and Objective Function for MIMO Model}
\subsubsection{MIMO Communication Model and CI constraint}
The original form of communication model that containing matrix convolution is not easy to handle. To that end, we convert matrix convolution to matrix multiplication by rewriting equation (\ref{eq:c_model_3}) as
\begin{equation}
\label{eq:c_model_4}
    {\mathbf{Y}_C}^{\top} = (\mathbf{G}\mathbf{U}_{\phi})^{\top}\overline{\mathbf{X}}_C{\mathbf{H}_C}^{\top} + {\mathbf{N}_C}^{\top},
\end{equation}
where
\begin{equation}
\label{eq:c_def_X}
    \overline{\mathbf{X}}_C = 
    \begin{bmatrix}
        \mathbf{x}_{C,1}^{\top} & \mathbf{0} & \cdots & \mathbf{0} \\
        \mathbf{x}_{C,2}^{\top} & \mathbf{x}_{C,1}^{\top} & \ddots & \vdots \\
        \vdots & \mathbf{x}_{C,2}^{\top} & \ddots & \mathbf{0} \\
        \mathbf{x}_{C,L_0}^{\top} & \vdots & \ddots & \mathbf{x}_{C,1}^{\top} \\
        \mathbf{0} & \mathbf{x}_{C,L_0}^{\top} & \ddots & \mathbf{x}_{C,2}^{\top} \\
        \vdots & \vdots & \ddots & \vdots \\
        \mathbf{0} & \mathbf{0} & \cdots & \mathbf{x}_{C,L_0}^{\top}
    \end{bmatrix}\in\mathbb{C}^{L_1\times NP}.
\end{equation}

\textit{Proposition 2:} By defining $\mathbf{E}_{p}=[\mathbf{0}_{L_0\times(p-1)},\mathbf{I}_{L_0},\mathbf{0}_{L_0\times(P-p)}]^{\top}$, we are able to rewrite equation (\ref{eq:c_model_4}) as
\begin{equation}
\begin{aligned}
    \text{vec}(\mathbf{Y}_C^{\top}) = \overline{\mathbf{H}}_C\text{vec}(\mathbf{S}^{\top}) + \text{vec}(\mathbf{N}_C^{\top}),
\end{aligned}
\end{equation}
where
\begin{equation}
\begin{aligned}
    \overline{\mathbf{H}}_C &= (\mathbf{H}_C\otimes(\mathbf{G}\mathbf{U}_{\phi})^{\top})
    \begin{bmatrix}
        \mathbf{I}_{N_t}\otimes(\mathbf{E}_{1}\mathbf{\Omega}_{\phi}) \\
        \mathbf{I}_{N_t}\otimes(\mathbf{E}_{2}\mathbf{\Omega}_{\phi}) \\
        \vdots \\
        \mathbf{I}_{N_t}\otimes(\mathbf{E}_{P}\mathbf{\Omega}_{\phi})
    \end{bmatrix},
\end{aligned}
\end{equation}
and $\text{vec}(\mathbf{N}_C^{\top})\thicksim\mathcal{CN}(\mathbf{0},\sigma_C^2\mathbf{I}_K\otimes\mathbf{\Lambda}_{\phi})$.

\textit{Proof:} \nocolor{See section Appendix B}.

CI constraint represents the constraint that pushes the received symbols away from their corresponding detection thresholds within the modulated-symbol constellation, thereby contributing positively to the overall useful signal power. According to \cite{masouros2015exploiting}, for any transmitted symbol $d$ and its corresponding \nocolor{noise-free} received symbol $y$, the CI constraint guarantees that
\begin{equation}
\label{eq:original_ci}
    \left| \Im\left\{d^* y\right\} \right| - \Re\left\{d^* y\right\}\tan\theta \leq (-\sqrt{\Gamma}\tan\theta)\sigma.
\end{equation}
where $\Gamma$ represents the requisite SINR at the receiver end, \nocolor{$\theta$ is related to the type of constellation and is $\pi/4$ for QPSK}, and $\sigma$ denotes the standard deviation of the corresponding noise imposed on $y$.

Suppose $\overline{\mathbf{H}}_C\text{vec}(\mathbf{S}^{\top})=[\mathbf{y}_{C,1}^{\top},\mathbf{y}_{C,2}^{\top},\cdots,\mathbf{y}_{C,K}^{\top}]^{\top}$, where $\mathbf{y}_{C,k}$ and $\mathbf{d}_k$ represent the symbol stream received by the $k$-th user in the absence of noise and the symbol stream intended for transmission to the $k$-th user, respectively. Define $\bm{\varsigma}=\sqrt{\mathrm{diag}(\sigma_C^2\mathbf{I}_K\otimes\mathbf{\Lambda}_{\phi})}$, following the inequality (\ref{eq:original_ci}), the CI constraint imposed on the $k$-th user can be expressed as
\begin{equation}
    \left| \Im\left\{\mathbf{d}_k^*\circ\mathbf{y}_{C,k}\right\} \right| - \Re\left\{\mathbf{d}_k^*\circ\mathbf{y}_{C,k}\right\}\tan\theta \leq (-\sqrt{\Gamma_k}\tan\theta)\bm{\varsigma},\forall k
\end{equation}
where \nocolor{$\circ$ denotes the Hadamard product, and} $\Gamma_k$ is the required SNR for the $k$-th user. Subsequently, we aim to consolidate the $k$ CI constraints and recast them into a single matrix inequality.

\textit{Proposition 3:} Define $\mathbf{\Gamma}=\mathrm{Diag}([\Gamma_1,\Gamma_2,\cdots,\Gamma_K]^{\top})$ and $\overline{\mathbf{D}}=\mathrm{Diag}(\mathbf{\text{vec}(\mathbf{D}^{\top})})$ ; then, the CI constraint for $k$ users can be formulated as
\begin{equation}
\label{eq:mimo_ci}
\begin{aligned}
    \left| \Im\left\{\overline{\mathbf{D}}^{*}\overline{\mathbf{H}}_C\text{vec}(\mathbf{S}^{\top})\right\} \right| - \Re\left\{\overline{\mathbf{D}}^{*}\overline{\mathbf{H}}_C\text{vec}(\mathbf{S}^{\top})\right\}\tan\theta \\
    \leq (-\sqrt{\mathbf{\Gamma}\otimes\mathbf{I}_L}\tan\theta)\bm{\varsigma}.
\end{aligned}
\end{equation}

\textit{Proof:} \nocolor{See section Appendix C}.

\subsubsection{MIMO Radar Model and MMSE for Sensing}
Similar to communication model, the radar model can also be expressed as
\begin{equation}
\label{eq:r_model_4}
    {\mathbf{Y}_R}^{\top} = \overline{\mathbf{X}}_R{\mathbf{H}_R}^{\top} + {\mathbf{N}_R}^{\top},
\end{equation}
where $\overline{\mathbf{X}}_R$ is defined in the same fashion with equation (\ref{eq:c_def_X}) as
\begin{equation}
\label{eq:r_def_X}
    \overline{\mathbf{X}}_R = [\mathbf{E}_{1}\mathbf{X}_R^{\top},\mathbf{E}_{2}\mathbf{X}_R^{\top},\cdots,\mathbf{E}_{P}\mathbf{X}_R^{\top}].
\end{equation}
Different from communication model, we extract $\text{vec}(\mathbf{H}_R^{\top})$ for the sake of derivation of MMSE. Thus we have
\begin{equation}
     \text{vec}(\mathbf{Y}_R^{\top}) = (\mathbf{I}_{N_r}\otimes\overline{\mathbf{X}}_R)\text{vec}(\mathbf{H}_R^{\top}).
\end{equation}
Let us assume $\text{vec}(\mathbf{H}_R^{\top})=\mathbf{h}\thicksim\mathcal{CN}(\mathbf{0},\sigma_H^2\mathbf{I})$, then according to \cite{kay1993fundamentals,tang2021constrained}, MMSE with respect to $\mathbf{h}$ can be written as
\begin{equation}
\begin{aligned}
    \mathrm{MMSE} &= \sigma_R^2\text{tr}\left(\left(\frac{\sigma_R^2}{\sigma_H^2}\mathbf{I}+(\mathbf{I}_{N_r}\otimes\overline{\mathbf{X}}_R)^H(\mathbf{I}_{N_r}\otimes\overline{\mathbf{X}}_R)\right)^{-1}\right) \\
    &= \sigma_R^2N_r\text{tr}\left(\left(\frac{\sigma_R^2}{\sigma_H^2}\mathbf{I}+\overline{\mathbf{X}}_R^H\overline{\mathbf{X}}_R\right)^{-1}\right) \\
    &= N_r\text{tr}(\sigma_H^2\mathbf{I} - \sigma_H^4\overline{\mathbf{X}}^H(\sigma_H^2\overline{\mathbf{X}}_R\overline{\mathbf{X}}_R^H + \sigma_R^2\mathbf{I})^{-1}\overline{\mathbf{X}}_R),
\end{aligned}
\end{equation}
where the MMSE estimator is expressed as
\begin{equation}
   \hat{\mathbf{h}}_\mathrm{MMSE} = \sigma_H^2\overline{\mathbf{X}}_R^H(\sigma_H^2\overline{\mathbf{X}}_R\overline{\mathbf{X}}_R^H + \sigma_R^2\mathbf{I})^{-1}\mathbf{y}. 
\end{equation}
\nocolor{Upon successful estimation of the TRM, it becomes feasible to determine the parameters of the sensing channel, utilizing the resultant TRM matrices.}

\subsubsection{MIMO Energy Constraint}
Due to the fact that our transmitted impulses are no longer orthogonal to each other, using $\Vert\mathbf{S}\Vert_F^2$ as the energy might not be suitable.

\textit{Proposition 4:} The energy of the transmitted waveform $x_n(t)$ at $n$-th antenna is given by
\begin{equation}
    \int\Vert x_n(t)\Vert^2 dt = \mathbf{s}_n^{H}\mathbf{\Phi}\mathbf{s}_n.
\end{equation}
where 
\begin{equation}
\begin{aligned}
    &\mathbf{\Phi} = \\
    &\begin{bmatrix}
        \phi(0) & \phi(-T) & \cdots & \phi(-(L-1)T) \\
        \phi(T) & \phi(0) & \cdots & \phi(-(L-2)T) \\
        \vdots & \vdots & \ddots & \vdots \\
        \phi((L-1)T) & \phi((L-2)T) & \cdots & \phi(0)
    \end{bmatrix}.
\end{aligned}
\end{equation}

\textit{Proof:} \nocolor{See section Appendix D}.

Therefore, the energy constraint under a given budget $E$ may be written as
\begin{equation}
    \sum_{n=1}^{N_t}\mathbf{s}_n^{H}\mathbf{\Phi}\mathbf{s}_n=\text{tr}(\mathbf{S}\mathbf{\Phi}\mathbf{S}^H)\leq E.
\end{equation}
or in the following quadratic form with respect to $\text{vec}(\mathbf{S}^{\top})$
\begin{equation}
    \text{vec}(\mathbf{S}^{\top})^H(\mathbf{I}_{N_t}\otimes\mathbf{\Phi})\text{vec}(\mathbf{S}^{\top})\leq E.
\end{equation}
\nocolor{Alternatively, one may also impose a per-antenna energy constraint, namely
\begin{equation}
    \mathbf{s}_n^{H}\mathbf{\Phi}\mathbf{s}_n=\text{tr}(\mathbf{S}\mathbf{\Phi}\mathbf{S}^H)\leq E_n,\;\forall n.
\end{equation}
We will compare the resultant MMSE performance under both energy constraints in numerical results.}

\section{FTN-ISAC Symbol-Level Precoding \nocolor{Design}}
\subsection{Problem Formulation}
Based on the discussion above, the precoding optimization problem for MIMO model can be expressed as
\begin{equation}
\label{opt:origin}
\begin{aligned}
    \underset{\mathbf{S}}{\min} &\;
    f(\mathbf{S})=\text{tr}\left(\left(\frac{\sigma_R^2}{\sigma_H^2}\mathbf{I}+\overline{\mathbf{X}}_R^H\overline{\mathbf{X}}_R\right)^{-1}\right) \\
    s.t. &\;
    \left| \Im\left\{\overline{\mathbf{D}}^{*}\overline{\mathbf{H}}_C\text{vec}(\mathbf{S}^{\top})\right\} \right| - \Re\left\{\overline{\mathbf{D}}^{*}\overline{\mathbf{H}}_C\text{vec}(\mathbf{S}^{\top})\right\}\tan\theta \\
    &\qquad\qquad\qquad\qquad\qquad\qquad
    \leq (-\sqrt{\mathbf{\Gamma}\otimes\mathbf{I}_L}\tan\theta)\bm{\varsigma}, \\
    &\; \text{vec}(\mathbf{S}^{\top})^H(\mathbf{I}_{N_t}\otimes\mathbf{\Phi})\text{vec}(\mathbf{S}^{\top})\leq E, \\
    &\;\nocolor{\mathbf{X}_R = \mathbf{S}\mathbf{\Omega}_{\varphi}^{\top}, \overline{\mathbf{X}}_R = [\mathbf{E}_{1}\mathbf{X}_R^{\top},\mathbf{E}_{2}\mathbf{X}_R^{\top},\cdots,\mathbf{E}_{P}\mathbf{X}_R^{\top}].}
\end{aligned}
\end{equation}
\nocolor{By formulating the problem above, we aim to} construct the pre-encoded symbols $\mathbf{S}$ corresponding to the given data matrix $\mathbf{D}$ intended for transmission, in such a manner that the MMSE pertinent to radar detection is reduced to its lowest possible value, whilst concurrently ensuring the CI constraint are satisfied in the context of a \nocolor{energy budget} $E$.

\subsection{Minorization Approach for FTN-ISAC-SLP}
It is important to acknowledge that the optimization problem (\ref{opt:origin}) is non-convex in nature. To tackle this challenge, we devise an optimization framework based on the minorization approach in this section.

Solving problem (\ref{opt:origin}) is equivalent to solving
\begin{equation}
\label{opt:origin}
\begin{aligned}
    \underset{\mathbf{S}}{\max} &\;
    f_m(\mathbf{S})=\text{tr}(\overline{\mathbf{X}}^H(\sigma_H^2\overline{\mathbf{X}}\overline{\mathbf{X}}^H + \sigma_R^2\mathbf{I})^{-1}\overline{\mathbf{X}}) \\
    s.t. &\;
    \left| \Im\left\{\overline{\mathbf{D}}^{*}\overline{\mathbf{H}}_C\text{vec}(\mathbf{S}^{\top})\right\} \right| - \Re\left\{\overline{\mathbf{D}}^{*}\overline{\mathbf{H}}_C\text{vec}(\mathbf{S}^{\top})\right\}\tan\theta \\
    &\qquad\qquad\qquad\qquad\qquad\qquad
    \leq (-\sqrt{\mathbf{\Gamma}\otimes\mathbf{I}_L}\tan\theta)\bm{\varsigma}, \\
    &\; \text{vec}(\mathbf{S}^{\top})^H(\mathbf{I}_{N_t}\otimes\mathbf{\Phi})\text{vec}(\mathbf{S}^{\top})\leq E.
\end{aligned}
\end{equation}
The fundamental concept of the proposed framework revolves around deriving a minorizer for $f_m(\mathbf{S})$. More specifically, the derived minorizers (denoted by $g_m(\mathbf{S}; \mathbf{S}_k)$) ought to satisfy the following conditions:
\begin{equation}
\begin{aligned}
    g_m(\mathbf{S}; \mathbf{S}_k)\leq f_m(\mathbf{S}), \;
    g_m(\mathbf{S}_k; \mathbf{S}_k)=f_m(\mathbf{S}_k),
\end{aligned}
\end{equation}

According to the work of \cite{tang2021constrained}, we can construct minorizer using the following inequality 
\begin{equation}
\begin{aligned}
    f_m(\mathbf{S})&=\text{tr}\left(\overline{\mathbf{X}}_R^{H}\left(\sigma_H^2\overline{\mathbf{X}}_R\overline{\mathbf{X}}_R^{H} + \sigma_R^2\mathbf{I}\right)^{-1}\overline{\mathbf{X}}_R\right) \\ &\geq
    2\Re\left\{\text{tr}(\mathbf{Q}_k^H\overline{\mathbf{X}}_R)\right\} - \text{tr}(\mathbf{T}_k(\sigma_H^2\overline{\mathbf{X}}_R\overline{\mathbf{X}}_R^H+\sigma_R^2\mathbf{I})) 
\end{aligned}
\end{equation}
where 
\begin{equation}
\begin{aligned}
    \mathbf{Q}_k=\sigma_H^4(\sigma_H^2\overline{\mathbf{X}}_{R,k}\overline{\mathbf{X}}_{R,k}^H+\sigma_R^2\mathbf{I})^{-1}\overline{\mathbf{X}}_{R,k}, 
    \mathbf{T}_k=\mathbf{Q}_k\mathbf{Q}_k^H/\sigma_H^4. 
\end{aligned}
\end{equation}

\textit{Proposition 5:} By defining 
\begin{equation}
\begin{aligned}
    \mathbf{E}_R
    = \begin{bmatrix}
        \mathbf{I}_N\otimes(\mathbf{E}_{1}\mathbf{\Omega}_{\varphi}) \\
        \mathbf{I}_N\otimes(\mathbf{E}_{2}\mathbf{\Omega}_{\varphi}) \\
        \vdots \\
        \mathbf{I}_N\otimes(\mathbf{E}_{P}\mathbf{\Omega}_{\varphi})
    \end{bmatrix},
\end{aligned}
\end{equation}
we are able to minorize $f_m(\overline{\mathbf{X}})$ by
\begin{equation}
    g_m(\mathbf{S};\mathbf{S}_k)=c_k - 2\Re\left\{\text{vec}(\mathbf{S}^{\top})^H\mathbf{b}_k\right\}-\text{vec}(\mathbf{S}^{\top})^H\mathbf{B}_k\text{vec}(\mathbf{S}^{\top}).
\end{equation}
where $c_k=-\text{tr}(\sigma_R^2\mathbf{T}_k)$,  and 
\begin{align}
    \mathbf{b}_k&=-\mathbf{E}_R^H\text{vec}(\mathbf{Q}_k), \label{eq:b}\\
    \mathbf{B}_k&=\mathbf{E}_R^H(\sigma_H^2\mathbf{I}\otimes\mathbf{T}_k)\mathbf{E}_R\succeq\mathbf{0}. \label{eq:B}
\end{align}

\textit{Proof:} \nocolor{See section Appendix E}.

Then we can express minorizing problem at $k+1$-th iteration as
\begin{equation}
\label{opt:minorization}
\begin{aligned}
    \underset{\mathbf{S}}{\min} &\;
    2\Re\left\{\text{vec}(\mathbf{S}^{\top})^H\mathbf{b}_k\right\}+\text{vec}(\mathbf{S}^{\top})^H\mathbf{B}_k\text{vec}(\mathbf{S}^{\top}) \\
    s.t. &\;
    \left| \Im\left\{\overline{\mathbf{D}}^{*}\overline{\mathbf{H}}_C\text{vec}(\mathbf{S}^{\top})\right\} \right| - \Re\left\{\overline{\mathbf{D}}^{*}\overline{\mathbf{H}}_C\text{vec}(\mathbf{S}^{\top})\right\}\tan\theta
    \\ &\qquad\qquad\qquad\qquad\qquad\qquad
    \leq (-\sqrt{\mathbf{\Gamma}\otimes\mathbf{I}_L}\tan\theta)\bm{\varsigma},\\
    &\; \text{vec}(\mathbf{S}^{\top})^H(\mathbf{I}_{N_t}\otimes\mathbf{\Phi})\text{vec}(\mathbf{S}^{\top})\leq E.
\end{aligned}
\end{equation}
Assume that a solution $\mathbf{S}_k$ has been obtained upon the completion of the $k$-th iteration. Subsequently, by solving problem (\ref{opt:minorization}) at the $(k+1)$-th iteration, an optimal solution $\mathbf{S}^{\star}$ is acquired. We can confidently assert that $f_m(\mathbf{S}^{\star})\geq g_m(\mathbf{S}^{\star};\mathbf{S}_k)\geq g_m(\mathbf{S}_k;\mathbf{S}_k)=f_m(\mathbf{S}_k)$, which implies that a superior solution for minimizing MMSE can be attained at the $(k+1)$-th iteration. This process can be iteratively repeated to continuously optimize the solution until that the convergence is achieved.

We are now ready to present Algorithm \ref{alg:minorization} to solve problem (\ref{opt:origin})  based on the discussion above. The detail concerning solving problem (\ref{opt:minorization}) is elaborated in section \ref{sec:efficient_qp}.
\begin{algorithm}
\caption{Minorization Method for Solving (\ref{opt:origin})}
\label{alg:minorization}
\begin{algorithmic}[1]
    \REQUIRE $N_t$, $L$, $\sigma_C^2$, $\sigma_R^2$, $\sigma_H^2$, $\mathbf{D}$, $\mathbf{H}_C$, $\mathbf{\Gamma}$, the execution threshold $\epsilon$ and the maximum iteration number $i_{\max}$.
    \ENSURE $\mathbf{S}^{\star}$
    \STATE {initialize $\mathbf{S}_0\in\mathcal{Q}$ by picking up $\mathbf{S}_{-1}$ randomly and solving problem (\ref{opt:minorization}), $k=0$.}
    \REPEAT{
        \STATE Calculate the $\mathbf{b}_k$ and $\mathbf{B}_k$ by equation (\ref{eq:b}) and (\ref{eq:B}).
        \STATE Solve problem (\ref{opt:minorization}) to obtain $\mathbf{S}^{\star}$.
        \STATE $k=k+1$.
    }
    \UNTIL {
        $\Vert \mathbf{S}_k-\mathbf{S}_{k-1}\Vert_F^2\leq\epsilon$ or $i=i_{\max}$.
    }
    \STATE $\mathbf{S}^{\star}=\mathbf{S}_k$
\end{algorithmic}
\end{algorithm}

\subsection{SCA Approach for FTN-ISAC-SLP}
In this section, we will develop another optimization scheme to solve problem (\ref{opt:origin}) by the idea of SCA. To proceed with the SCA algorithm, we approximate the objective function by its first-order Taylor expansion near a given point $\mathbf{S}_k$ (and hence $\overline{\mathbf{X}}_{R,k}$) as
\begin{equation}
    f(\mathbf{S}) \approx f(\mathbf{S}_k) + \Re\left\{\text{vec}\left(\frac{\partial f}{\partial \overline{\mathbf{X}}_R}\right)^{H}\text{vec}\left(\overline{\mathbf{X}}_R-\overline{\mathbf{X}}_{R,k}\right)\right\}
\end{equation}
where
\begin{equation}
\label{eq:gradient}
    \frac{\partial f}{\partial \overline{\mathbf{X}}_R} = -2\overline{\mathbf{X}}_{R,k}\left(\overline{\mathbf{X}}_{R,k}^{H}\overline{\mathbf{X}}_{R,k}\right)^{-1}\left(\overline{\mathbf{X}}_{R,k}^{H}\overline{\mathbf{X}}_{R,k}\right)^{-1}
\end{equation}
stands for the gradient at the point $\mathbf{S}_k$. By using the fact that $\text{vec}(\overline{\mathbf{X}}_R)=\mathbf{E}_R\text{vec}(\mathbf{S}^{\top})$ and define 
\begin{equation}
\label{eq:t_R}
    \mathbf{t}_{R,k}^{H} = \text{vec}\left(\frac{\partial f}{\partial \overline{\mathbf{X}}_R}\right)^{H}\mathbf{E}_R,
\end{equation}
we are able to approximate $f(\mathbf{S})$ around $\mathbf{S}_k$ by
\begin{equation}
\begin{aligned}
    g_l(\mathbf{S};\mathbf{S}_k) \approx \;&\Re\left\{\mathbf{t}_{R,k}^{H}\text{vec}(\mathbf{S}^{\top})\right\} \\
    &+ f(\mathbf{S}_k) - \Re\left\{\text{vec}\left(\frac{\partial f}{\partial \overline{\mathbf{X}}_R}\right)^{H}\text{vec}\left(\overline{\mathbf{X}}_{R,k}\right)\right\}.
\end{aligned}
\end{equation}
Then we proceed to solve the following sub-problem (\ref{opt:SCA}) in $(k+1)$-th iteration.
\begin{equation}
\label{opt:SCA}
\begin{aligned}
    \underset{\mathbf{S}}{\mathrm{min}} &\;
    \Re\left\{\mathbf{t}_{R,i}^{H}\text{vec}(\mathbf{S}^{\top})\right\} \\
    s.t. &\;
    \left| \Im\left\{\overline{\mathbf{D}}^{*}\overline{\mathbf{H}}_C\text{vec}(\mathbf{S}^{\top})\right\} \right| - \Re\left\{\overline{\mathbf{D}}^{*}\overline{\mathbf{H}}_C\text{vec}(\mathbf{S}^{\top})\right\}\tan\theta \\
    &\qquad\qquad\qquad\qquad\qquad\qquad
    \leq (-\sqrt{\mathbf{\Gamma}\otimes\mathbf{I}_L}\tan\theta)\bm{\varsigma}, \\
    &\; \text{vec}(\mathbf{S}^{\top})^H(\mathbf{I}_{N_t}\otimes\mathbf{\Phi})\text{vec}(\mathbf{S}^{\top})\leq E.
\end{aligned}
\end{equation}
Suppose that a solution $\mathbf{S}_k$ has been procured at the $k$-th iteration. By subsequently solving problem (\ref{opt:SCA}) during the $(k+1)$-th iteration, an optimal solution $\mathbf{S}^{\star}$ is obtained. When $\mathbf{S}^{\star}$ is in close proximity to $\mathbf{S}_k$ and the SCA approximation holds, it follows that $f(\mathbf{S}^{\star})\leq g_l(\mathbf{S}^{\star};\mathbf{S}_k)\leq g_l(\mathbf{S}_k;\mathbf{S}_k)=f(\mathbf{S}_k)$. Albeit $\mathbf{S}^{\star}$ is not necessarily adjacent to $\mathbf{S}_k$, the difference $\mathbf{S}^{\star}-\mathbf{S}_k$ provides a decent direction for the optimization of $f(\mathbf{S})$. By iteratively taking small steps along the direction of $\mathbf{S}^{\star}-\mathbf{S}_k$, it is possible to successively obtain superior solutions that minimize MMSE prior to reaching convergence.

With a properly chosen step size $t\in[0,1]$, one may get the $(k+1)$-th iteration point as
\begin{equation}
    \mathbf{S}_{k+1}=\mathbf{S}_k+t(\mathbf{S}^{\star}-\mathbf{S}_k)=(1-t)\mathbf{S}_k+t\mathbf{S}^{\star}.
\end{equation}
Since $\mathbf{S}_k,\mathbf{S}^{\star} \in \mathcal{Q}$ by the definition of convexity, we have $\mathbf{S}_{k+1} \in \mathcal{Q}$, which is a feasible solution to problem (\ref{opt:origin}).

We are now ready to present Algorithm \ref{alg:SCA} to solve problem (\ref{opt:origin})  based on the discussion above. The detail concerning solving problem (\ref{opt:minorization}) is elaborated in section \ref{sec:efficient_qp}.
\begin{algorithm}
\caption{SCA Method for Solving (\ref{opt:origin})}
\label{alg:SCA}
\begin{algorithmic}[1]
    \REQUIRE $N_t$, $L$, $\sigma_C^2$, $\sigma_R^2$, $\sigma_H^2$, $\mathbf{D}$, $\mathbf{H}_C$, $\mathbf{\Gamma}$, the execution threshold $\epsilon$ and the maximum iteration number $i_{\max}$.
    \ENSURE $\mathbf{S}^{\star}$
    \STATE {Initialize $\mathbf{S}_0\in\mathcal{Q}$ by picking up $\mathbf{S}_{-1}$ randomly and solving problem (\ref{opt:SCA}), $k=0$.}
    \REPEAT{
        \STATE Calculate the $\mathbf{t}_{R,k}$ by equation (\ref{eq:gradient}) and (\ref{eq:t_R}).
        \STATE Solve problem (\ref{opt:SCA}) to obtain $\mathbf{S}^{\star}$.
        \STATE {
            Update the solution by
            $\mathbf{S}_{k+1} = \mathbf{S}_k + t\left(\mathbf{S}^{\star}-\mathbf{S}_k\right)$,
            where $t$ is determined by using the exact line search.
        }
        \STATE $k=k+1$.
    }
    \UNTIL {
        $\Vert \mathbf{S}_k-\mathbf{S}_{k-1}\Vert_F^2\leq\epsilon$ or $i=i_{\max}$.
    }
    \STATE $\mathbf{S}^{\star}=\mathbf{S}_k$
\end{algorithmic}
\end{algorithm}

\subsection{Efficient Algorithm for Solving Sub-problems (\ref{opt:minorization}) and (\ref{opt:SCA})}
\label{sec:efficient_qp}
Notice that both problems (\ref{opt:minorization}) and (\ref{opt:SCA}) can be written in the form of 
\begin{equation}
\label{opt:energy_qp}
\begin{aligned}
    \underset{\mathbf{S}}{\min} &\;
    2\Re\left\{\text{vec}(\mathbf{S}^{\top})^H\mathbf{a}_k\right\}+\text{vec}(\mathbf{S}^{\top})^H\mathbf{A}_k\text{vec}(\mathbf{S}^{\top}) \\
    s.t. &\;
    \left| \Im\left\{\overline{\mathbf{D}}^{*}\overline{\mathbf{H}}_C\text{vec}(\mathbf{S}^{\top})\right\} \right| - \Re\left\{\overline{\mathbf{D}}^{*}\overline{\mathbf{H}}_C\text{vec}(\mathbf{S}^{\top})\right\}\tan\theta
    \\ &\qquad\qquad\qquad\qquad\qquad\qquad
    \leq (-\sqrt{\mathbf{\Gamma}\otimes\mathbf{I}_L}\tan\theta)\bm{\varsigma},\\
    &\; \text{vec}(\mathbf{S}^{\top})^H(\mathbf{I}_{N_t}\otimes\mathbf{\Phi})\text{vec}(\mathbf{S}^{\top})\leq E.
\end{aligned}
\end{equation}
which is a QCQP with linear inequality constraint and quadratic energy constraint. Specifically, for problem (\ref{opt:SCA}), $\mathbf{A}_k=\mathbf{0}$. Note that numerous efficient methods have been developed to solve real QP with linear inequality constraints, such as problem (\ref{opt:qp}), for both small-scale and large-scale problems.
\begin{equation}
\label{opt:qp}
\begin{aligned}
    &\underset{\mathbf{x}}{\mathrm{min}} \;
    \mathbf{x}^{\top}\mathbf{A}\mathbf{x} + 2\mathbf{x}^{\top}\mathbf{a} \\
    &s.t.\;
    \mathbf{B}\mathbf{x} \leq \mathbf{b}.
\end{aligned}
\end{equation}
Therefore, in order to enhance the efficiency of our algorithm, which involves solving QCQP problems (\ref{opt:minorization}) and (\ref{opt:SCA}), we aim to develop an algorithm that can transform the process of solving QCQP into solving QP problems with only linear inequality constraints.

By letting $\mathbf{P}_{\Re} = \Re\left\{\overline{\mathbf{D}}^{*}\overline{\mathbf{H}}_C\right\}$, $\mathbf{P}_{\Im} = \Im\left\{\overline{\mathbf{D}}^{*}\overline{\mathbf{H}}_C\right\}$, $\mathbf{s}_{\Re} = \Re\left\{\text{vec}(\mathbf{S}^{\top})\right\}$ and $\mathbf{s}_{\Im} = \Im\left\{\text{vec}(\mathbf{S}^{\top})\right\}$, the inequalities (\ref{eq:mimo_ci}) can be decomposed to
\begin{equation}
\begin{aligned}
    (\mathbf{P}_{\Im}-\mathbf{P}_{\Re}\tan\theta)\mathbf{s}_{\Re} + (\mathbf{P}_{\Re}+\mathbf{P}_{\Im}\tan\theta)\mathbf{s}_{\Im}& \\
    \leq (-\sqrt{\mathbf{\Gamma}\otimes\mathbf{I}_L}\tan\theta)\bm{\varsigma}&, \\
    (-\mathbf{P}_{\Im}-\mathbf{P}_{\Re}\tan\theta)\mathbf{s}_{\Re} + (-\mathbf{P}_{\Re}+\mathbf{P}_{\Im}\tan\theta)\mathbf{s}_{\Im}& \\
    \leq (-\sqrt{\mathbf{\Gamma}\otimes\mathbf{I}_L}\tan\theta)\bm{\varsigma}&,
\end{aligned}
\end{equation}
or in single matrix inequality form
\begin{equation}
\label{eq:matrix_ci}
\begin{aligned}
    \begin{bmatrix}
        \mathbf{P}_{\Im}-\mathbf{P}_{\Re}\tan\theta & \mathbf{P}_{\Re}+\mathbf{P}_{\Im}\tan\theta \\
        -\mathbf{P}_{\Im}-\mathbf{P}_{\Re}\tan\theta & -\mathbf{P}_{\Re}+\mathbf{P}_{\Im}\tan\theta
    \end{bmatrix}
    \begin{bmatrix}
        \mathbf{s}_{\Re}\\
        \mathbf{s}_{\Im}
    \end{bmatrix}& \\
    \leq 
    \begin{bmatrix}
        (-\sqrt{\mathbf{\Gamma}\otimes\mathbf{I}_L}\tan\theta)\bm{\varsigma}\\
        (-\sqrt{\mathbf{\Gamma}\otimes\mathbf{I}_L}\tan\theta)\bm{\varsigma}
    \end{bmatrix}&.
\end{aligned}
\end{equation}
Taking
\begin{equation}
\begin{aligned}
    &\mathbf{\Psi} =
    \begin{bmatrix}
        \mathbf{P}_{\Im}-\mathbf{P}_{\Re}\tan\theta & \mathbf{P}_{\Re}+\mathbf{P}_{\Im}\tan\theta \\
        -\mathbf{P}_{\Im}-\mathbf{P}_{\Re}\tan\theta & -\mathbf{P}_{\Re}+\mathbf{P}_{\Im}\tan\theta
    \end{bmatrix}, \\
    &\hat{\mathbf{s}} = 
    \begin{bmatrix}
        \mathbf{s}_{\Re}\\
        \mathbf{s}_{\Im}
    \end{bmatrix},
    \bm{\gamma} = 
    \begin{bmatrix}
        (-\sqrt{\mathbf{\Gamma}\otimes\mathbf{I}_L}\tan\theta)\bm{\varsigma}\\
        (-\sqrt{\mathbf{\Gamma}\otimes\mathbf{I}_L}\tan\theta)\bm{\varsigma}
    \end{bmatrix} \\
    &\hat{\mathbf{A}}_{k} =
    \begin{bmatrix}
        \Re\left\{\mathbf{B}_{k}\right\} & -\Im\left\{\mathbf{B}_{k}\right\} \\
        \Im\left\{\mathbf{B}_{k}\right\} & \Re\left\{\mathbf{B}_{k}\right\}
    \end{bmatrix}, 
    \hat{\mathbf{a}}_{k} =
    \begin{bmatrix}
        \Re\left\{\mathbf{b}_{k}\right\} \\
        -\Im\left\{\mathbf{b}_{k}\right\}
    \end{bmatrix}, \\
    &\mathbf{\Upsilon} =
    \begin{bmatrix}
        \mathbf{I}_{N_t}\otimes\mathbf{\Phi} & \mathbf{0} \\
        \mathbf{0} & \mathbf{I}_{N_t}\otimes\mathbf{\Phi}
    \end{bmatrix}, 
\end{aligned}
\end{equation}
we can rewrite problem (\ref{opt:minorization}) as
\begin{equation}
\label{opt:real_origin}
\begin{aligned}
    &\underset{\hat{\mathbf{s}}}{\min} \;
    \hat{\mathbf{s}}^{\top}\hat{\mathbf{A}}_{k}\hat{\mathbf{s}} + 2\hat{\mathbf{s}}^{\top}\hat{\mathbf{a}}_{k} \\
    &s.t.\;
    \hat{\mathbf{s}}^{\top}\mathbf{\Upsilon}\hat{\mathbf{s}} \leq E, \; \mathbf{\Psi}\hat{\mathbf{s}} \leq \bm{\gamma}.
\end{aligned}
\end{equation}

We then try to remove the energy constraint in (\ref{opt:energy_qp}) by introducing the penalty factor $\rho\geq 0$ and the penalty problem (\ref{opt:penalty_qp}).
\begin{equation}
\label{opt:penalty_qp}
\begin{aligned}
    \mathcal{K}(\rho):\; &\underset{\hat{\mathbf{s}}}{\min} \;
    \hat{\mathbf{s}}^{\top}(\hat{\mathbf{A}}_{k}+\rho\mathbf{\Upsilon})\hat{\mathbf{s}} + 2\hat{\mathbf{s}}^{\top}\hat{\mathbf{a}}_{k} \\
    &s.t.\;
    \mathbf{\Psi}\hat{\mathbf{s}} \leq \bm{\gamma}
\end{aligned}
\end{equation}

\nocolor{The key idea here is to introduce} a regularization term, $\rho\hat{\mathbf{s}}^{\top}\mathbf{\Upsilon}\hat{\mathbf{s}}$, into the objective function to ensure that the energy term $\hat{\mathbf{s}}^{\top}\mathbf{\Upsilon}\hat{\mathbf{s}}$ does not become excessively large. However, it is important to note that if $\rho$ is too large, the optimal solution of problem (\ref{opt:penalty_qp}) would primarily minimize $\hat{\mathbf{s}}^{\top}\mathbf{\Upsilon}\hat{\mathbf{s}}$, consequently leading to inadequate reduction of the original objective function $ \hat{\mathbf{s}}^{\top}\hat{\mathbf{B}}_k\hat{\mathbf{s}} + 2\hat{\mathbf{s}}^{\top}\hat{\mathbf{b}}_k$. Conversely, if $\rho$ is too small, the optimal solution would not sufficiently consider the term $\hat{\mathbf{s}}^{\top}\mathbf{\Upsilon}\hat{\mathbf{s}}$, potentially resulting in violation of the constraint $\hat{\mathbf{s}}^{\top}\mathbf{\Upsilon}\hat{\mathbf{s}} \leq E$. To address this issue, we propose a binary penalty search (BPS) algorithm, which aims to identify an appropriate value for $\rho$ and effectively solve problem (\ref{opt:real_origin}).

Suppose we now have two penalty factors, $\rho_r > \rho_l > 0$, such that solving $\mathcal{K}(\rho_r)$ results in a solution $\hat{\mathbf{s}_r}$ that satisfies the energy constraint $\hat{\mathbf{s}_r}^{\top}\mathbf{\Upsilon}\hat{\mathbf{s}_r} \leq E$, while solving $\mathcal{K}(\rho_l)$ yields a solution $\hat{\mathbf{s}_l}$ that violates the energy constraint. We can infer that there may exist a $\rho$ in the interval $(\rho_l, \rho_r)$, for which solving $\mathcal{K}(\rho)$ provides a solution that more effectively minimizes the objective function than solving $\mathcal{K}(\rho_r)$, while still adhering to the energy constraint.

By realizing the fact above, we set the penalty factor $\rho = (\rho_r + \rho_l) / 2$, which is the mid-point of the interval $(\rho_l, \rho_r)$. If solving $\mathcal{K}(\rho)$ yields a new solution that violates the energy constraint, we set $\rho_l = \rho$. Otherwise, we set $\rho_r = \rho$. This approach progressively narrows the search range. By iteratively applying this process, we eventually reach a sufficiently small search range where $\rho_r \approx \rho_l$. The most appropriate penalty factor would then be $\rho_r$. Consequently, the Binary Penalty Search (BPS) algorithm can be presented in Algorithm \ref{alg:binary_penalty_search}.

\textit{Proposition 6:} The solution derived by this algorithm is the optimal solution of the problem (\ref{opt:real_origin}).

\textit{Proof:} \nocolor{See section Appendix F}.

\begin{algorithm}[htbp]
\caption{Binary Penalty Search}
\label{alg:binary_penalty_search}
\begin{algorithmic}[1]
    \REQUIRE $\hat{\mathbf{A}}_{k}$, $\hat{\mathbf{a}}_{k}$, $E$, $\mathbf{\Psi}$, $\mathbf{\Upsilon}$, $\bm{\gamma}$, execution threshold $\epsilon$.
    \ENSURE optimal solution $\hat{\mathbf{s}}_{\star}$
    \STATE Initialize $\rho_l=0$, $\rho_r=\rho_{max}$.
    \REPEAT
        \STATE $\rho=(\rho_l+\rho_r)/2$;
        \STATE Solve problem (\ref{opt:penalty_qp}) by active-set method to obtain $\hat{\mathbf{s}}_{\star}$.
        \IF {$\hat{\mathbf{s}}_{\star}^{\top}\mathbf{\Upsilon}\hat{\mathbf{s}}_{\star}\leq E$}
            \STATE $\rho_r=\rho$;
        \ELSE
            \STATE $\rho_l=\rho$;
        \ENDIF
    \UNTIL{$\rho_r-\rho_l\leq\epsilon$}
\end{algorithmic}
\end{algorithm}

\textit{Comparison of Time Complexity between Primal Dual Interior Point and BPS Algorithm:} The computational complexity of the Primal-Dual Interior Point Method (PDIP) algorithm for addressing problem (\ref{opt:energy_qp}) typically falls within the bounds of $\text{O}((4NL+1)^{3.5}\log(1/\epsilon))$ to $\text{O}((4NL+1)^4\log(1/\epsilon))$. This signifies a marginal elevation in computational complexity in contrast to problem (\ref{opt:penalty_qp}), which exhibits a time complexity of $\text{O}((4NL+1)^{3.5}\log(1/\epsilon))$.

When employing the BPS approach to solve problem (\ref{opt:energy_qp}), the procedure necessitates the resolution of $\log(\rho_{max})$ instances of problem (\ref{opt:penalty_qp}). Consequently, the algorithmic time complexity for the BPS approach escalates to $\text{O}(\log(\rho_{max}/\epsilon)(4NL)^{3.5}\log(1/\epsilon))$.
In essence, this indicates that the BPS algorithm's computational complexity surpasses that of the PDIP algorithm when the constraints imposed by the energy limit are less stringent, signified by a comparably diminished $\rho_{max}$.

Regarding problem (\ref{opt:penalty_qp}), various efficacious algorithms for QP can be adopted to accommodate a diversity of situations. For example, when operating within a relatively small-scale problem space, the active-set strategy can be implemented to lower the practical computational complexity effectively.

\section{Numerical Results}
In this section, we provide numerical results to verify the superiority of the proposed FTN-ISAC-SLP approaches. Without loss of generality, we consider an ISAC BS that is equipped with $N_r = 8$ antennas for its receiver. The noise variances are set as $\sigma^2_C = \sigma^2_R = 0\;\mathrm{dBm}$. The quantity of $\bm{\mu}_\mathbf{h}$ has minimal impact on the optimization discussed in this paper and is therefore set to $\mathbf{0}$. The variance of TRM fluctuations is set as $\sigma^2_H=20\;\mathrm{dBm}$, with each element of $\mathbf{H}_R$ drawn from $\mathcal{CN}(0,\sigma_H^2)$. Symbol duration $T_0$ is set to $1\;\mathrm{ms}$. Each element of $\mathbf{H}_C$ independently is drawn from $\mathcal{CN}(0,\sigma_C^2)$. Without loss of generality, all the communication users are imposed with the same worst-case QoS, i.e., $\Gamma_k=\Gamma,\forall k$.

\begin{figure}[th]
    \centering
    \includegraphics[width=0.9\linewidth]{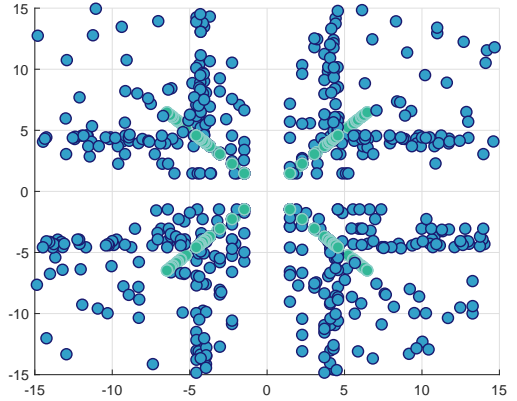}
    \caption{Constellation plot of the received symbols without noise. Here we take $\tau=0.9$. The green dots are the nominal constellation points.}
    \label{fig:constellation}
\end{figure}

Fig. \ref{fig:constellation} presents the constellation plot of the received symbols of the FTN-ISAC-SLP system before noise imposition. Generally, the symbols are distanced from the detection thresholds, namely, the x and y axes. However, some symbols are observed to be closer to the thresholds. This occurrence can be attributed to the noise $\text{vec}(\mathbf{N}_C^{\top})\thicksim\mathcal{CN}(\mathbf{0},\sigma_C^2\mathbf{I}K\otimes\mathbf{\Lambda}{\phi})$, which is imposed on the symbols and possesses varying variances. Consequently, certain symbols are more susceptible to noise interference, while others remain unaffected. The system adapts by allocating more power to the received symbols subjected to noise with higher variance.

\begin{figure}[t]
    \centering
    \begin{minipage}{0.9\linewidth}
        \centering
        \includegraphics[width=0.9\linewidth]{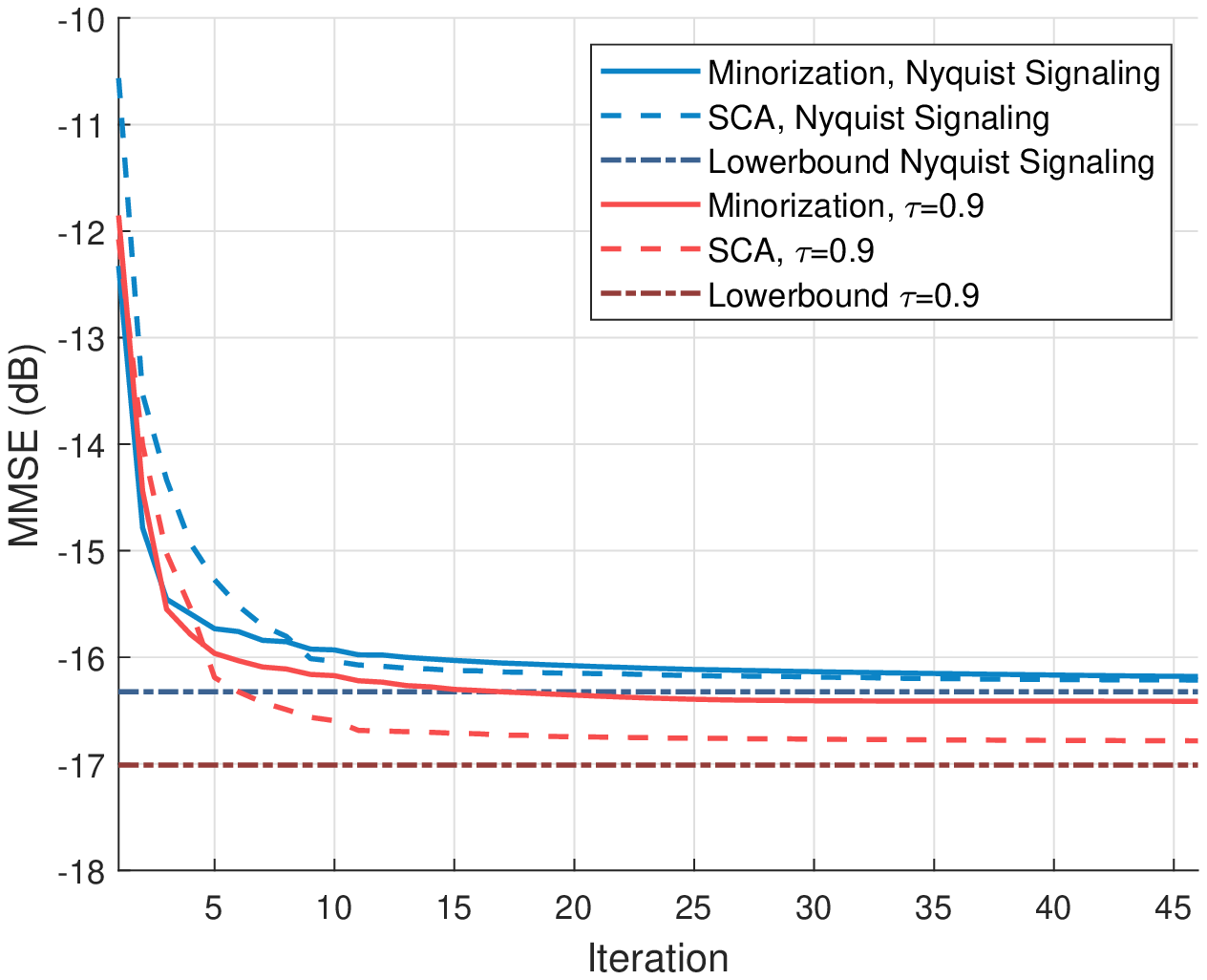}
        \caption{MMSE versus minorization/SCA iteration in case of $N_t=3$, $K=2$, $L=15$, $P=3$, $Q=3$, $\Gamma=15\;\mathrm{dBm}$, $E=30\;\mathrm{dBm}$.}
        \label{fig:MIMO_MMSEvsIter_MinvsLin}
    \end{minipage}
    \begin{minipage}{0.9\linewidth}
        \centering
        \includegraphics[width=0.9\linewidth]{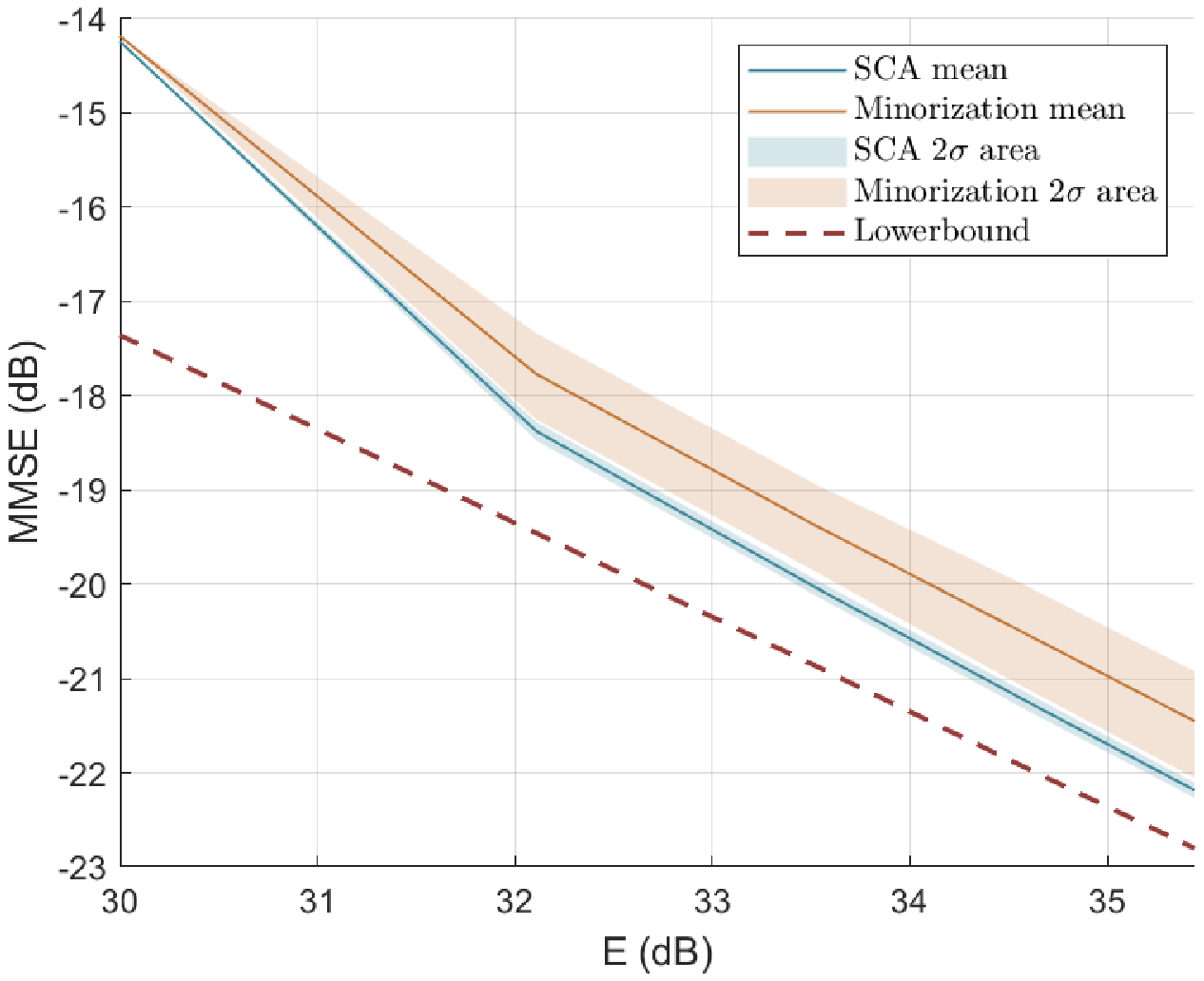}
        \caption{MMSE versus $E$ in case of $N_t=3$, $K=3$, $L=20$, $P=3$, $Q=3$, $\Gamma = 15\;\mathrm{dBm}$, $\tau=0.9$.}
        \label{fig:MIMO_MMSEvsE}
    \end{minipage}
\end{figure}

In Fig. \ref{fig:MIMO_MMSEvsIter_MinvsLin}, we compare the convergence behavior of Algorithm \ref{alg:minorization} (minorization approach) and Algorithm \ref{alg:SCA} (SCA approach) by illustrating how MMSE changes with the number of iterations. It is observed that the minorization approach reaches a better sensing performance in the early iterations, while the SCA approach ultimately achieves better results after more iterations.
The slower convergence of the SCA approach can be due to the fact that the SCA approximation requires the solution to take only small steps along the optimization direction to ensure that the approximation remains valid. This leads to more iterations being needed for the SCA approach to converge to the ultimate solution. However, it is worth noting that despite its slower convergence, the SCA approach eventually results in better sensing performance, which could be advantageous in scenarios where the final performance is of higher importance than the speed of convergence.

In Fig. \ref{fig:MIMO_MMSEvsE}, we compare the performance of the minorization and SCA approaches under different energy budgets $E$ and fixed communication and sensing conditions, using $1000$ initial data points for each method. Subsequently, we compute the mean and standard deviation of the MMSE generated by distinct initial points. The $2\sigma$ region denotes the area encompassing points situated within a distance of twice the standard deviation from the mean.
As the energy budget $E$ increases, both methods exhibit better sensing performance and approach the lower bound of MMSE. This is expected, as a higher energy budget allows for more flexibility in satisfying both the communication and sensing requirements.
It is observed that the minorization approach exhibits a larger fluctuation in the results compared to the SCA approach, as indicated by the wider $2\sigma$ region. This suggests that the SCA approach might be more robust and consistent in terms of its performance across different initial data points.

\begin{figure}[t]
    \centering
    \begin{minipage}{0.9\linewidth}
        \centering
        \includegraphics[width=0.9\linewidth]{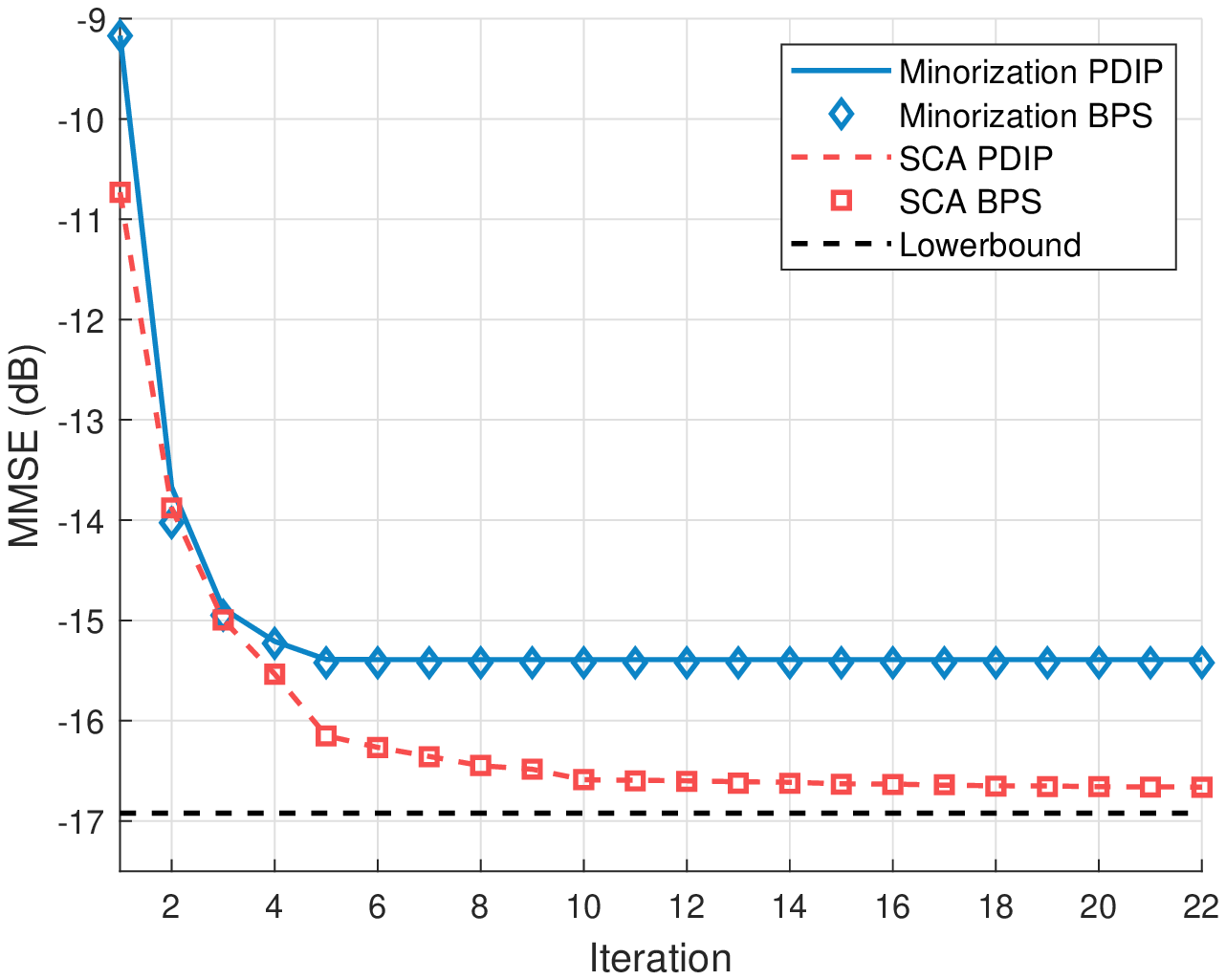}
        \caption{MMSE versus minorization/SCA iteration in case of $N=3$, $K=2$, $L=15$, $P=3$, $Q=3$, $\Gamma=15\;\mathrm{dBm}$, $E=30\;\mathrm{dBm}$, $\tau=0.9$.}
        \label{fig:MIMO_MMSEvsIter}
    \end{minipage}
    \begin{minipage}{0.9\linewidth}
        \centering
        \includegraphics[width=0.9\linewidth]{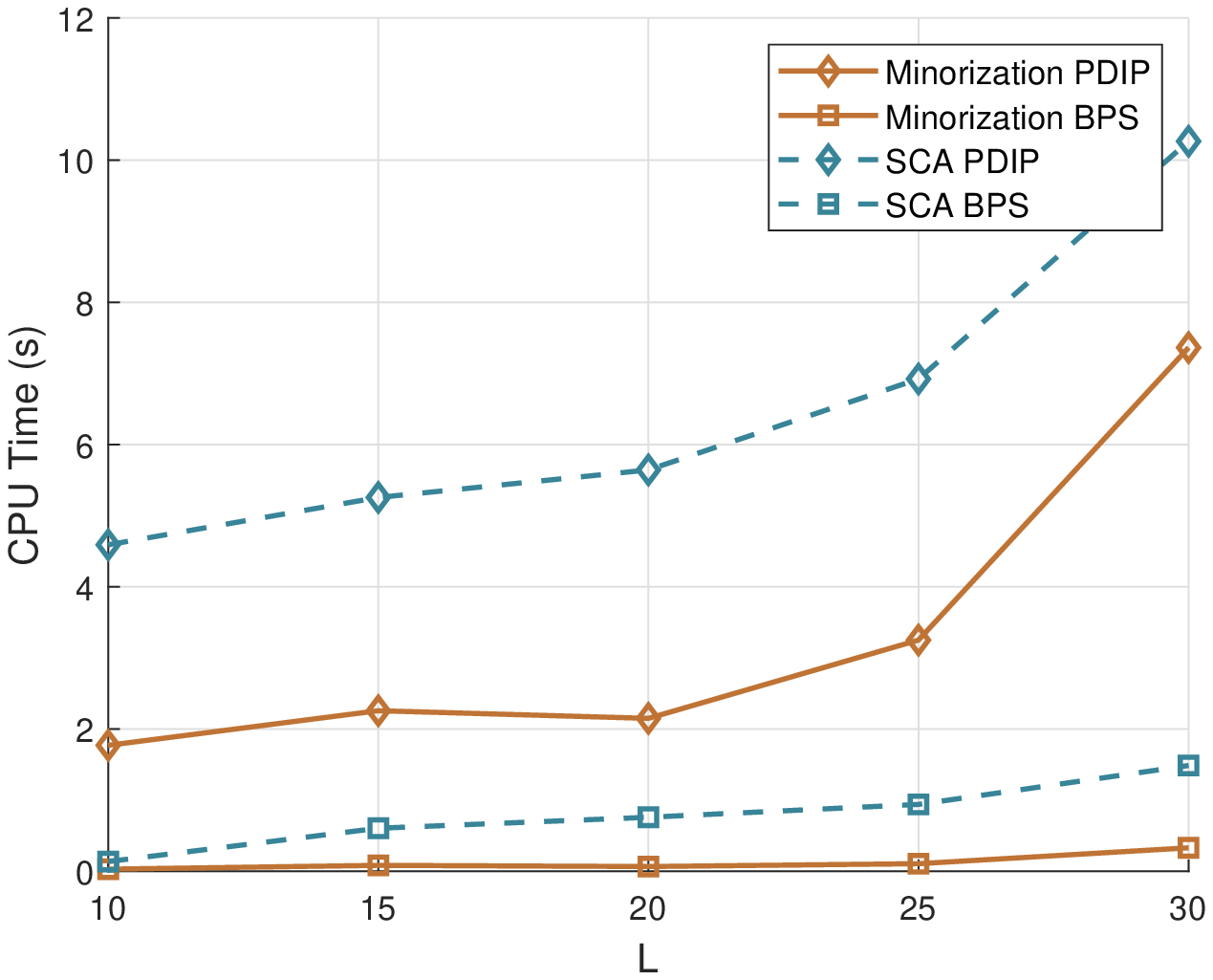}
        \caption{CPU Time versus $L$ in case of $N_t=3$, $K=2$, $P=3$, $Q=3$, $\Gamma=15\;\mathrm{dBm}$, $E=35\;\mathrm{dBm}$.}
        \label{fig:MIMO_TimevsL}
    \end{minipage}
    \begin{minipage}{0.9\linewidth}
        \centering
        \includegraphics[width=0.9\linewidth]{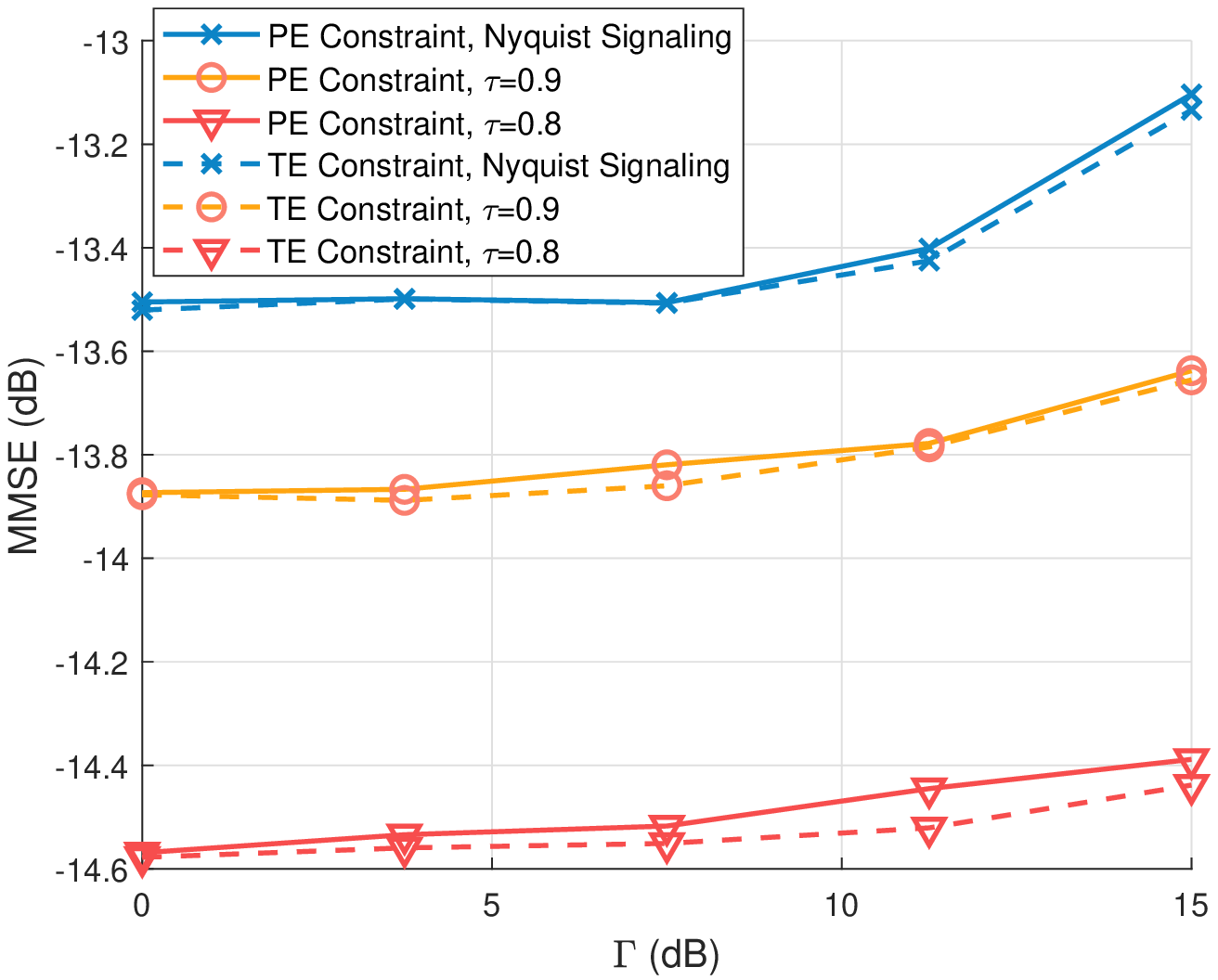}
        \caption{MMSE versus $\Gamma$ comparison between total energy (TE) constraint and per-antenna (PE) constraint in case of $N_t=4$, $K=3$, $L=15$, $P=3$, $Q=3$, $E = 30\;\mathrm{dBm}$.}
        \label{fig:perantenna}
    \end{minipage}
\end{figure}

In Fig. \ref{fig:MIMO_MMSEvsIter}, we compare the results of solving the minorization problems using the BPS algorithm and the Primal-Dual Interior Point (PDIP) algorithm directly. The purpose of this comparison is to evaluate the effectiveness of the BPS algorithm in solving the minorization problems, as compared to a well-established optimization algorithm like PDIP.
The figure shows that the BPS algorithm and the PDIP algorithm provide essentially the same results at each iteration of the algorithms. Moreover, as proved in the later section of the study, the BPS algorithm yields the optimal solution for the energy-constrained QP problem akin to the PDIP algorithm.

Fig. \ref{fig:MIMO_TimevsL} delineates the relationship between the Central Processing Unit (CPU) execution time and frame length $L$. The iterations are curtailed once the MMSE reaches a threshold of $-15;\text{dBm}$. As anticipated, an augmentation in $L$ corresponds to an expansion in the problem size, leading to a heightened demand for CPU time. The BPS exhibits a more efficient performance in terms of CPU time when juxtaposed with the PDIP method. Moreover, the minorization approach manifests superior convergence speed towards a reasonably low MMSE threshold compared to the SCA approach.

Fig. \ref{fig:perantenna} delineates the comparative analysis between the per-antenna and total energy constraints. For the sake of simplicity, we set each antenna's energy budget as $E_n=E/N_t$. The results indicate negligible discrepancy between the two constraints. Generally, the MMSE, when applied to the per-antenna energy constraint, is marginally higher than its counterpart under the total energy constraint.

\begin{figure}[t]
    \centering
    \begin{minipage}{0.9\linewidth}
        \centering
        \includegraphics[width=0.9\linewidth]{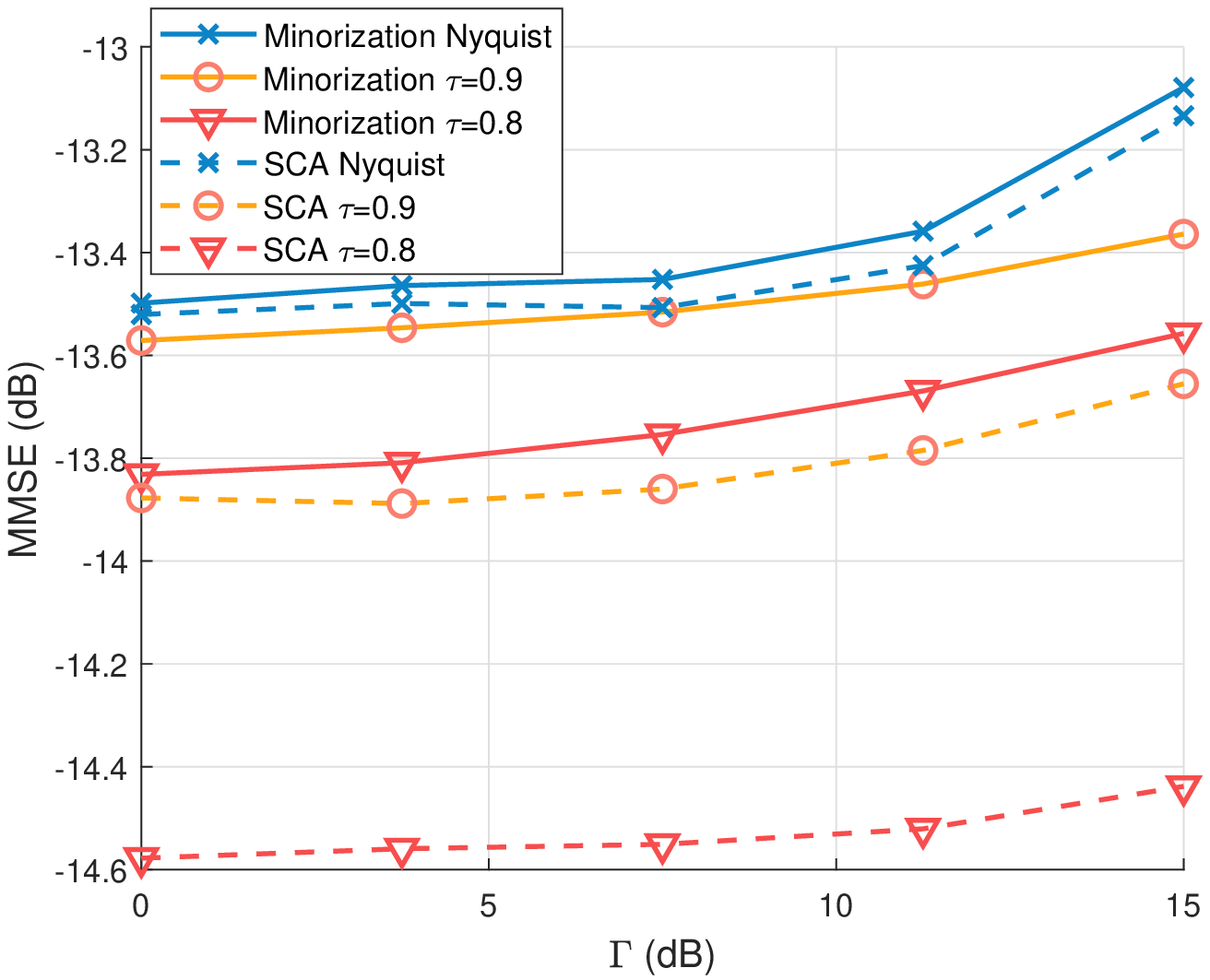}
        \caption{MMSE versus $\Gamma$ in case of $N_t=4$, $K=3$, $L=15$, $P=3$, $Q=3$, $E = 30\;\mathrm{dBm}$.}
        \label{fig:MIMO_MMSEvsSINR}
    \end{minipage}
    \begin{minipage}{0.9\linewidth}
        \centering
        \includegraphics[width=0.9\linewidth]{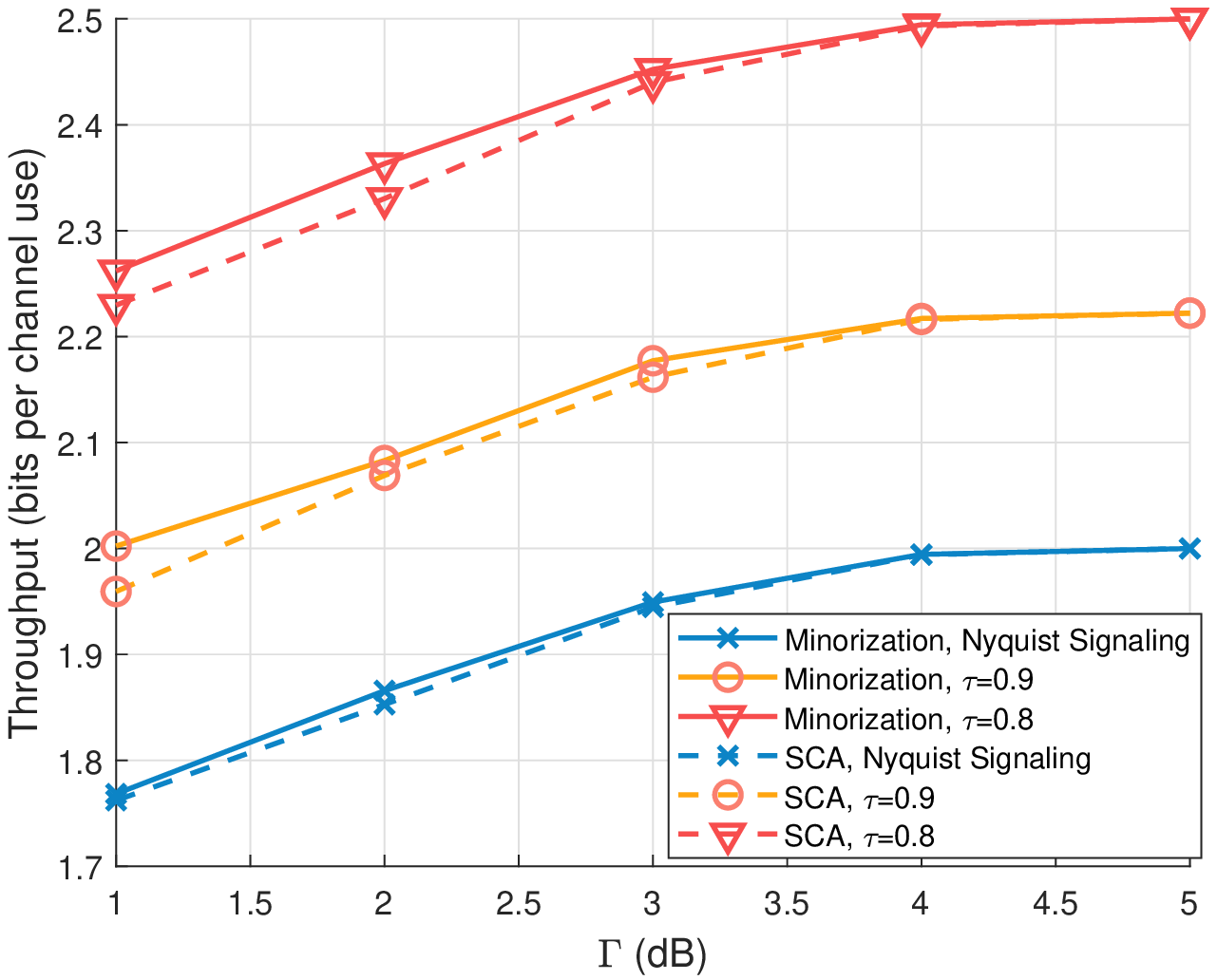}
        \caption{Throughput versus $\Gamma$ in case of $N_t=4$, $K=3$, $L=15$, $P=3$, $Q=3$, $E = 30\;\mathrm{dBm}$.}
        \label{fig:MIMO_TPvsSINR}
    \end{minipage}
    \begin{minipage}{0.9\linewidth}
        \centering
        \includegraphics[width=0.9\linewidth]{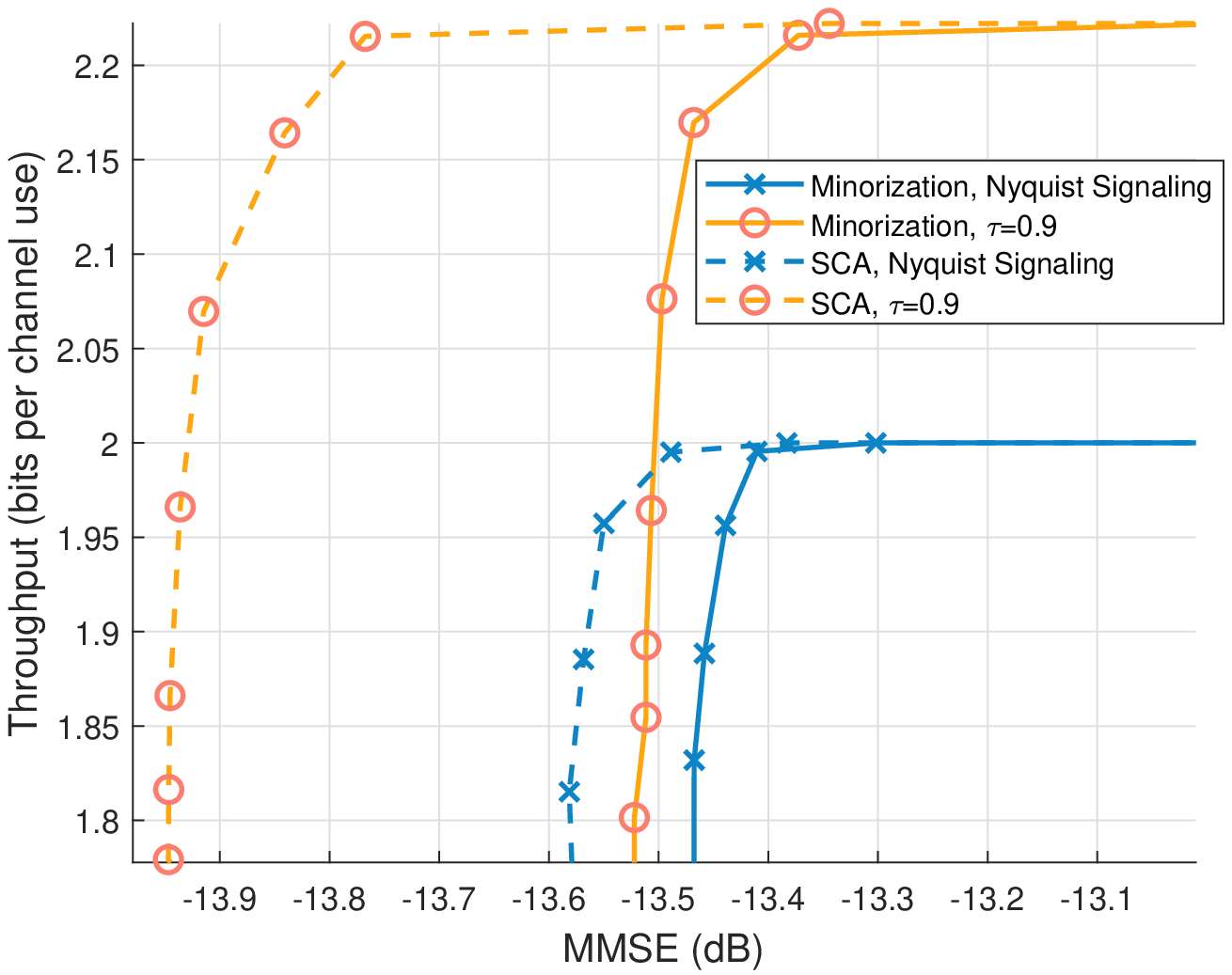}
        \caption{Throughput versus MMSE in case of $N_t=4$, $K=3$, $L=15$, $P=3$, $Q=3$, $E = 30\;\mathrm{dBm}$.}
        \label{fig:MIMO_TPvsMMSE}
    \end{minipage}
\end{figure}

\begin{figure}[t]
    \centering
    \begin{minipage}{0.9\linewidth}
        \centering
        \includegraphics[width=0.9\linewidth]{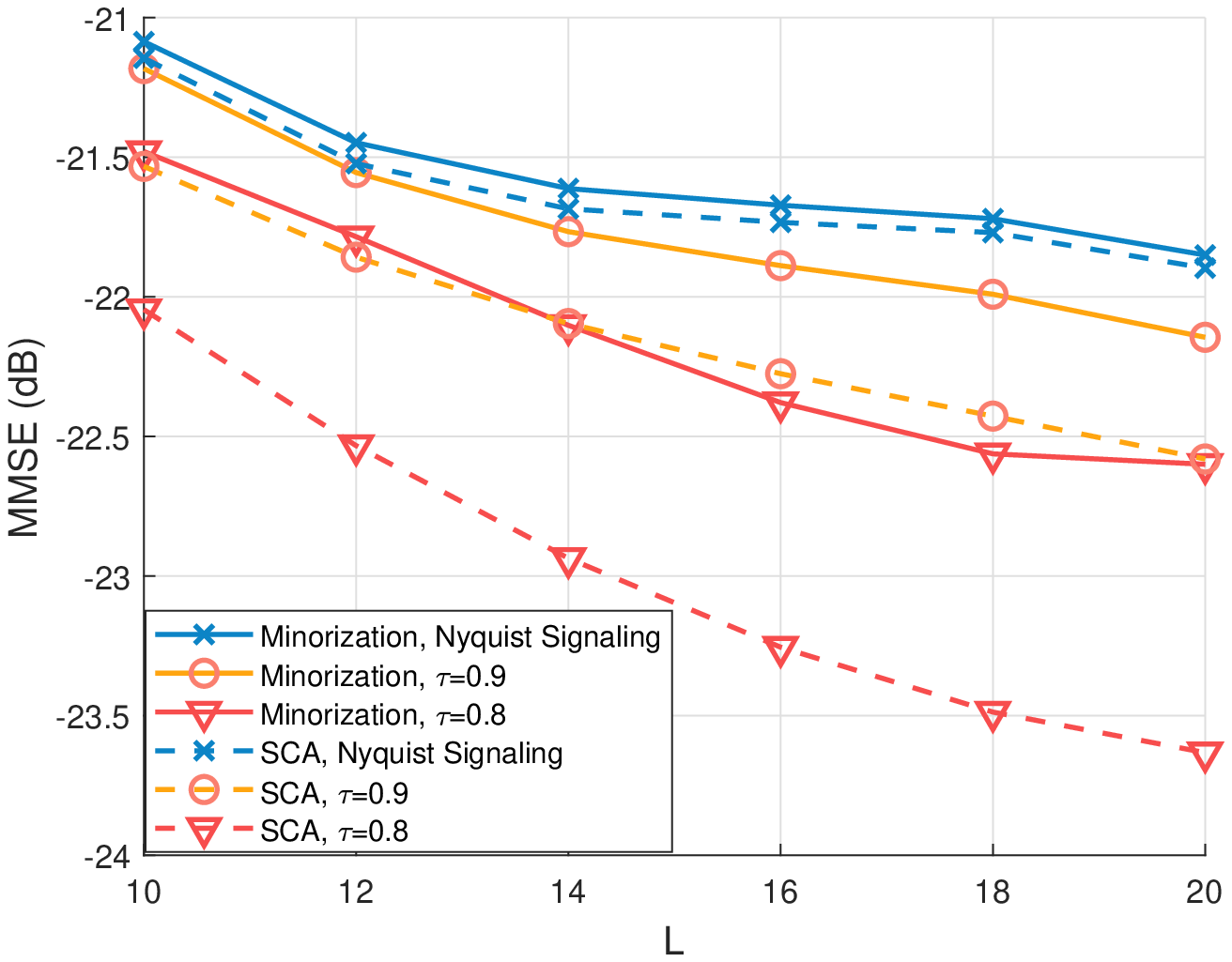}
        \caption{MMSE versus $L$ in case of $N_t=4$, $K=3$, $\Gamma=15\;\mathrm{dBm}$, $P=3$, $Q=3$, $E = 35\;\mathrm{dBm}$.}
        \label{fig:MIMO_MMSEvsL}
    \end{minipage}
    \begin{minipage}{0.9\linewidth}
        \centering
        \includegraphics[width=0.9\linewidth]{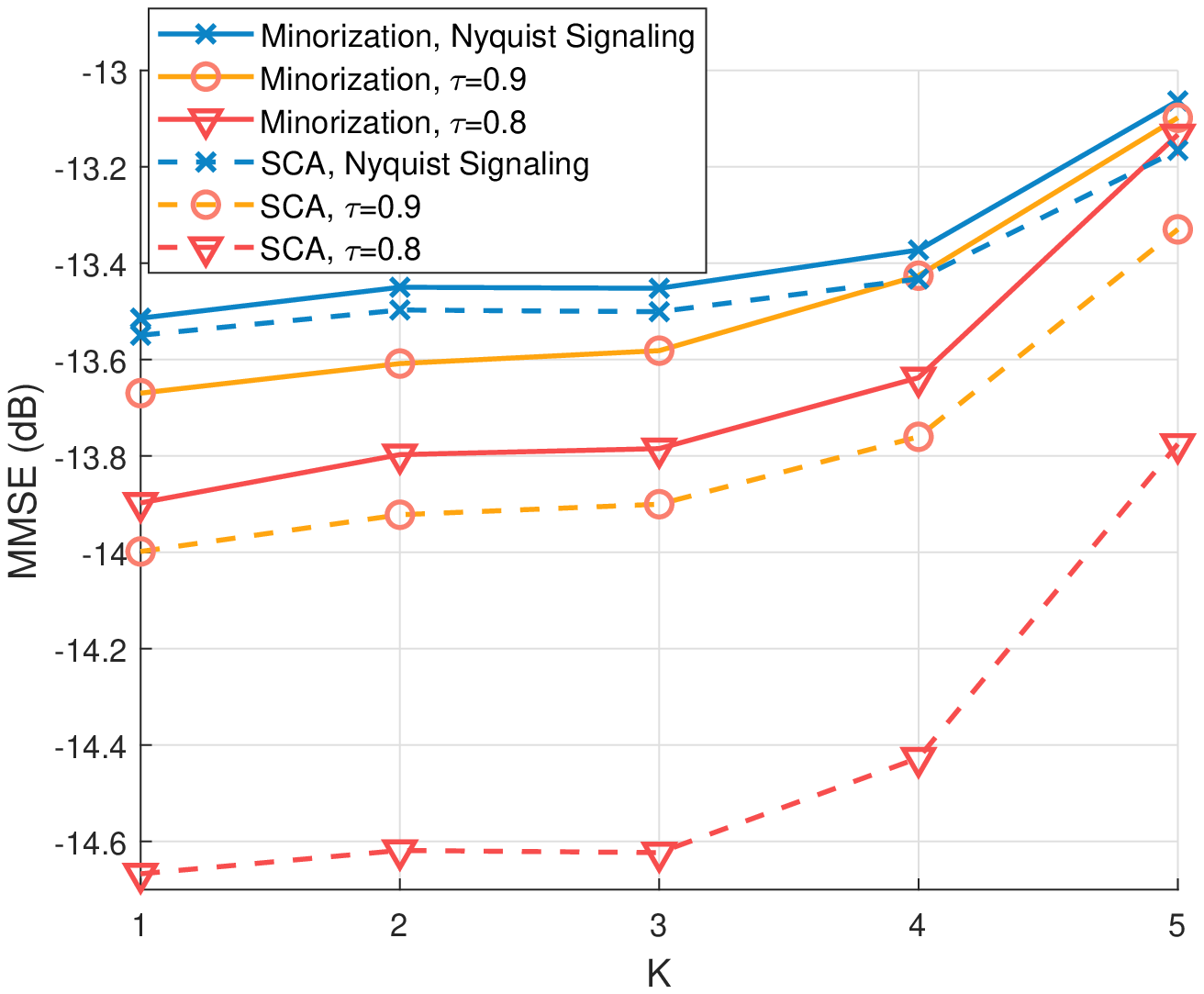}
        \caption{MMSE versus $K$ in case of $N_t=5$, $L=20$, $\Gamma=15\;\mathrm{dBm}$, $P=3$, $Q=3$, $E = 32\;\mathrm{dBm}$.}
        \label{fig:MIMO_MMSEvsK}
    \end{minipage}
\end{figure}

In Fig. \ref{fig:MIMO_MMSEvsSINR}, the impact of SNR threshold for communication users on the radar estimation MMSE is depicted. It can be observed that with an elevated communications SNR, the estimation performance deteriorates, signifying an intrinsic tradeoff between communication and sensing performance.
Concurrently, it is observed that as $\tau$ increases, the MMSE exhibits a decline. This phenomenon can be ascribed to the fact that interference in FTN signaling is harnessed to generate a positive impact on sensing performance. The smaller the value of $\tau$, the greater the extent to which one pulse can contribute its energy to adjacent pulses. In communication, such interference may lead to challenges in recovering the correct constellation. However, in sensing, a larger energy contribution often corresponds to enhanced sensing performance.

In Fig. \ref{fig:MIMO_TPvsSINR}, we demonstrate the impact of the SNR threshold for communication on communication throughput. \nocolor{Inspired by the computation of the bit error rate (BER) in \cite{vitthaladevuni2001ber}, we develop a way to calculate throughput from constellation that guarantee fair comparison under FTN signaling. Let's denote the symbol stream received sans noise as $\mathbf{y}$. In the context of QPSK, the probability of an individual symbol $y \in \mathbf{y}$ being received with an incorrect in-phase sign, under Gaussian noise, is represented as $p_1=Q(|\Re(y)|/\sqrt{\sigma_n^2/2})$. Concurrently, the probability for an incorrect quadrature sign is given by $p_2=Q(|\Im(y)|/\sqrt{\sigma_n^2/2})$, where $\sigma_n^2$ signifies the noise variance imposed on the symbol $y$ and $Q(\cdot)$ denotes the Q-function.
Accordingly, the probability for the correct transmission of the two bits encapsulated within this symbol is calculated as $(1-p_1)(1-p_2)$, and the probability for the correct transmission of a single bit is derived as $(1-p_1)p_2+p_1(1-p_2)$. 
The expected number of successfully transmitted bits is quantified as $2(1-p_1)(1-p_2)+(1-p_1)p_2+p_1(1-p_2)$, which is also $2(1-P_e)$ where $P_e$ is the calculation of BER in \cite{vitthaladevuni2001ber}. It is important to note that under FTN signaling, this symbol occupies only $\tau T_0$ time. To ensure a fair comparison, we normalize the expected number of successfully transmitted bits by $\tau$, which furnishes us with the formula for computing the throughput from a given constellation:
$$\frac{1}{KL}\sum_{y\in\mathbf{y}}\left(2(1-p_1)(1-p_2)+(1-p_1)p_2+p_1(1-p_2)\right)/\tau.$$}
It is observed that as the SNR threshold for communication increases, the communication throughput increases, which is essentially a result of received symbols being pushed further from the detection thresholds in the constellation plot.
At the same time, it is observed that as $\tau$ increases, the throughput increases, which is due to the fact that the same amount of symbols are transmitted in less time. This highlights the effectiveness of FTN signaling in achieving higher throughput for communication, while maintaining a balance between communication and sensing performance.

Fig. \ref{fig:MIMO_TPvsMMSE} illustrates the relationship between radar estimation MMSE and throughput. It can be observed that as MMSE increases, the throughput also increases. This indicates an inherent tradeoff between communication and sensing performance. When MMSE is sufficiently large, the throughput converges to the maximum transmission rate, signifying a bit error rate of zero. In this scenario, the system primarily emphasizes communication performance. Furthermore, it is noted that when throughput diminishes, the MMSE converges to a specific value, representing the lower bound of MMSE when the system is solely focused on sensing performance.

In Fig. \ref{fig:MIMO_MMSEvsL}, the influence of frame length $L$ on radar estimation performance is illustrated. It is observed that as the frame length increases, the estimation performance improves, even though communication performance constraints become more stringent. Additionally, in the case of SCA with $\tau=0.8$, the MMSE displays a convergence trend as $L$ increases, indicating that the sensing performance is limited by the energy constraint.

In Fig. \ref{fig:MIMO_MMSEvsK}, we demonstrate the impact of the communication user number $K$ on radar estimation performance. Unlike increasing the frame length $L$, which also increases the number of symbols to be transmitted, the estimation performance worsens as $K$ increases. This is because when we increase $L$ by $1$, we only tighten the communication constraint on $K$ more symbols. However, when we increase $K$ by $1$, we tighten the communication constraint on $L$ more symbols. In the two cases illustrated by Fig. \ref{fig:MIMO_MMSEvsL} and \ref{fig:MIMO_MMSEvsK}, $L$ in Fig. \ref{fig:MIMO_MMSEvsK} is significantly larger than $K$ in Fig. \ref{fig:MIMO_MMSEvsL}, resulting in different trends of MMSE.

\section{Conclusion}
In conclusion, this paper introduces a novel wideband FTN-ISAC-SLP precoding technique for MIMO DFRC systems, adeptly combining FTN signaling and SLP to enhance sensing and communication performance across temporal and spatial dimensions. To tackle the complex non-convex waveform design problem, we develop two algorithm frameworks based on minorization and SCA methods, transforming the problem into solvable QCQP sub-problems. Additionally, we propose a computationally efficient BPS method to solve these sub-problems. Extensive simulation results validate the effectiveness of the proposed FTN-ISAC-SLP design in both radar sensing and multi-user communication performance, contributing to the ongoing development of efficient and robust integrated sensing and communication technologies.

\section*{Appendix A: Proof of Proposition 1}
\label{app:prop1}
For any $i,j$ that $1\leq i,j\leq L$, we have
\begin{equation}
\begin{aligned}
    &\mathbb{E}[\eta_k(iT)\eta(jT)_k^*] \\
    &=\mathbb{E}\left[\int\mathbf{n}_c(t)\varphi^{*}(t-iT)dt\int\mathbf{n}_c^*(t)\varphi(t-jT)dt\right] \\
    &=\mathbb{E}\left[\int\int\mathbf{n}_c(t_1)\mathbf{n}_c(t_2)^{*}\varphi^{*}(t_1-iT)\varphi(t_2-jT)dt_1dt_2\right] \\
    &=\int\int\mathbb{E}[\mathbf{n}_c(t_1)\mathbf{n}_c(t_2)^{*}]\varphi^{*}(t_1-iT)\varphi(t_2-jT)dt_1dt_2 \\
    &=\int\int\sigma_C^2\delta(t_1-t_2)\varphi^{*}(t_1-iT)\varphi(t_2-jT)dt_1dt_2 \\
    &=\int\sigma_C^2\varphi^{*}(t-iT)\varphi(t-jT)dt \\
    &=\int\sigma_C^2\varphi^{*}(t)\varphi(t-(i-j)T)dt \\
    &=\sigma_C^2\phi((i-j)T)=\sigma_C^2\mathbf{\Phi}_{i,j}.
\end{aligned}
\end{equation}
Thus $\mathbb{E}[\bm{\eta}_k\bm{\eta}_k^H]=\sigma_C^2\mathbf{\Phi}$.

\section*{Appendix B: Proof of Proposition 2}
\label{app:prop2}
we can express $\overline{\mathbf{X}}_C$ as
\begin{equation}
    \overline{\mathbf{X}}_C = [\mathbf{E}_{1}\mathbf{X}_C^{\top},\mathbf{E}_{2}\mathbf{X}_C^{\top},\cdots,\mathbf{E}_{P}\mathbf{X}_C^{\top}],
\end{equation}
which yields a way to vectorize $\overline{\mathbf{X}}_C$ with respect to $\mathrm{vec}(\mathbf{S}^{\top})$
\begin{equation}
\begin{aligned}
    \mathrm{vec}(\overline{\mathbf{X}}_C) &= 
    \begin{bmatrix}
        \mathrm{vec}(\mathbf{E}_{1}\mathbf{X}_C^{\top}) \\
        \mathrm{vec}(\mathbf{E}_{2}\mathbf{X}_C^{\top}) \\
        \vdots \\
        \mathrm{vec}(\mathbf{E}_{P}\mathbf{X}_C^{\top})
    \end{bmatrix} = 
    \begin{bmatrix}
        \mathbf{I}_{N_t}\otimes\mathbf{E}_{1} \\
        \mathbf{I}_{N_t}\otimes\mathbf{E}_{2} \\
        \vdots \\
        \mathbf{I}_{N_t}\otimes\mathbf{E}_{P}
    \end{bmatrix}\mathrm{vec}(\mathbf{X}_C^{\top}) \\&= 
    \begin{bmatrix}
        \mathbf{I}_{N_t}\otimes(\mathbf{E}_{1}\mathbf{\Omega}_{\phi}) \\
        \mathbf{I}_{N_t}\otimes(\mathbf{E}_{2}\mathbf{\Omega}_{\phi}) \\
        \vdots \\
        \mathbf{I}_{N_t}\otimes(\mathbf{E}_{P}\mathbf{\Omega}_{\phi})
    \end{bmatrix}\mathrm{vec}(\mathbf{S}^{\top}).
\end{aligned}
\end{equation}
By using the fact that $\text{vec}(\mathbf{A}\mathbf{X}\mathbf{B})=(\mathbf{B}^{\top}\otimes\mathbf{A})\text{vec}(\mathbf{X})$, we have
\begin{equation}
\begin{aligned}
    \text{vec}(\mathbf{Y}_C^{\top}) = &(\mathbf{H}_C\otimes(\mathbf{G}\mathbf{U}_{\phi})^{\top})\text{vec}(\overline{\mathbf{X}}_C) + \text{vec}(\mathbf{N}_C^{\top}) \\
    = &\overline{\mathbf{H}}_C\text{vec}(\mathbf{S}^{\top}) + \text{vec}(\mathbf{N}_C^{\top}),
\end{aligned}
\end{equation}

\section*{Appendix C: Proof of Proposition 3}
\label{app:prop3}
Define $\mathbf{\Gamma}=\mathrm{Diag}([\Gamma_1,\Gamma_2,\cdots,\Gamma_K])$, $\overline{\mathbf{D}}=\mathrm{Diag}(\mathbf{\text{vec}(\mathbf{D}^{\top})})$ and $\bm{\varsigma}=\sqrt{\mathrm{diag}(\sigma_C^2\mathbf{I}_K\otimes\mathbf{\Lambda}_{\phi})}$; then, the CI constraint for $k$ users can be formulated as
\begin{equation}
\label{eq:mimo_ci}
\begin{aligned}
    \left| \Im\left\{\overline{\mathbf{D}}^{*}\overline{\mathbf{H}}_C\text{vec}(\mathbf{S}^{\top})\right\} \right| - \Re\left\{\overline{\mathbf{D}}^{*}\overline{\mathbf{H}}_C\text{vec}(\mathbf{S}^{\top})\right\}\tan\theta \\
    \leq (-\sqrt{\mathbf{\Gamma}\otimes\mathbf{I}_L}\tan\theta)\bm{\varsigma}.
\end{aligned}
\end{equation}
We first stack the $k$ linear inequalities
\begin{equation}
    \left| \Im\left\{\mathbf{d}_k^*\circ\mathbf{y}_{C,k}\right\} \right| - \Re\left\{\mathbf{d}_k^*\circ\mathbf{y}_{C,k}\right\}\tan\theta \leq (-\sqrt{\Gamma_k}\tan\theta)\bm{\sigma}
\end{equation}
in a column to form a united equality:
\begin{equation}
\label{eq:proof3_1}
\begin{aligned}
    &\left| \Im\left\{
    \begin{bmatrix}
        \mathbf{d}_1 \\
        \mathbf{d}_2 \\
        \vdots \\
        \mathbf{d}_k
    \end{bmatrix}^*
    \circ
    \begin{bmatrix}
        \mathbf{y}_{C,1} \\
        \mathbf{y}_{C,2} \\
        \vdots \\
        \mathbf{y}_{C,k}
    \end{bmatrix}\right\} \right| - 
    \Re\left\{
    \begin{bmatrix}
        \mathbf{d}_1 \\
        \mathbf{d}_2 \\
        \vdots \\
        \mathbf{d}_k
    \end{bmatrix}^*
    \circ
    \begin{bmatrix}
        \mathbf{y}_{C,1} \\
        \mathbf{y}_{C,2} \\
        \vdots \\
        \mathbf{y}_{C,k}
    \end{bmatrix}\right\}\tan\theta \\
    &=\left|\Im\left\{\text{vec}(\mathbf{D}^{\top})\circ\overline{\mathbf{H}}_C\text{vec}(\mathbf{S}^{\top}) \right\}\right| \\&- \Re\left\{\text{vec}(\mathbf{D}^{\top})\circ\overline{\mathbf{H}}_C\text{vec}(\mathbf{S}^{\top}) \right\}\tan\theta \leq
    \begin{bmatrix}
        (-\sqrt{\Gamma_1}\tan\theta)\bm{\varsigma} \\
        (-\sqrt{\Gamma_2}\tan\theta)\bm{\varsigma} \\
        \vdots \\
        (-\sqrt{\Gamma_k}\tan\theta)\bm{\varsigma}
    \end{bmatrix}.
\end{aligned}
\end{equation}
Notice that for Hardmard product of two vectors we have $\mathbf{a}\circ\mathbf{b}=\text{Diag}(\mathbf{a})\mathbf{b}$. Thus we can rewrite (\ref{eq:proof3_1}) as
\begin{equation}
\begin{aligned}
    &\left|\Im\left\{\text{Diag}(\text{vec}(\mathbf{D}^{\top}))\overline{\mathbf{H}}_C\text{vec}(\mathbf{S}^{\top}) \right\}\right| \\ &\qquad\qquad- \Re\left\{\text{Diag}(\text{vec}(\mathbf{D}^{\top}))\overline{\mathbf{H}}_C\text{vec}(\mathbf{S}^{\top}) \right\}\tan\theta \\
    &=\left| \Im\left\{\overline{\mathbf{D}}^{*}\overline{\mathbf{H}}_C\text{vec}(\mathbf{S}^{\top})\right\} \right| - \Re\left\{\overline{\mathbf{D}}^{*}\overline{\mathbf{H}}_C\text{vec}(\mathbf{S}^{\top})\right\}\tan\theta \\
    &\leq
    \begin{bmatrix}
        (-\sqrt{\Gamma_1}\tan\theta)\bm{\varsigma} \\
        (-\sqrt{\Gamma_2}\tan\theta)\bm{\varsigma} \\
        \vdots \\
        (-\sqrt{\Gamma_k}\tan\theta)\bm{\varsigma}
    \end{bmatrix}
    =(-\sqrt{\mathbf{\Gamma}\otimes\mathbf{I}_L}\tan\theta)\bm{\varsigma}.
\end{aligned}
\end{equation}

\section*{Appendix D: Proof of Proposition 4}
\label{app:prop4}
\begin{equation}
\begin{aligned}
    &\int\Vert x_n(t)\Vert^2 dt = \int\left\Vert\sum_{i=1}^{L}\varphi(t-(i-1)T)s_{n,i}\right\Vert^2 dt \\
    &=\int\left(\sum_{i=0}^{L-1}\varphi(t-iT)s_{n,i+1}\right)\left(\sum_{i=0}^{L-1}\varphi(t-iT)^*s_{n,i+1}^*\right) dt \\
    &=\int\sum_{i=1}^{L}\sum_{j=1}^{L}\left(\varphi(t-(i-1)T)\varphi(t-(j-1)T)^*s_{n,i}s_{n,j}^*\right) dt \\
    &=\sum_{i=1}^{L}\sum_{j=1}^{L}\left(\int\varphi(t-(i-1)T)\varphi(t-(j-1)T)^*dt\right)s_{n,i}s_{n,j}^* dt\\
    &=\sum_{i=1}^{L}\sum_{j=1}^{L}\left(\int\varphi(t)\varphi(t-(i-j)T)^*dt\right)s_{n,i}s_{n,j}^* dt\\
    &=\sum_{i=1}^{L}\sum_{j=1}^{L}\phi((i-j)T)s_{n,i}s_{n,j}^* = \mathbf{s}_n^{H}\mathbf{\Phi}\mathbf{s}_n.
\end{aligned}
\end{equation}

\section*{Appendix E: Proof of Proposition 5}
\label{app:prop5}
Similar to the vectorization of $\overline{\mathbf{X}}_C$, we are able to vectorize $\overline{\mathbf{X}}$ as
\begin{equation}
\begin{aligned}
    \text{vec}(\overline{\mathbf{X}}_R) &= \text{vec}([\mathbf{E}_1\mathbf{X}_R^{\top},\mathbf{E}_2\mathbf{X}_R^{\top},\cdots,\mathbf{E}_P\mathbf{X}_R^{\top}])\\
    &= \begin{bmatrix}
        \mathbf{I}_N\otimes(\mathbf{E}_{1}\mathbf{\Omega}_{\varphi}) \\
        \mathbf{I}_N\otimes(\mathbf{E}_{2}\mathbf{\Omega}_{\varphi}) \\
        \vdots \\
        \mathbf{I}_N\otimes(\mathbf{E}_{P}\mathbf{\Omega}_{\varphi})
    \end{bmatrix}\text{vec}(\mathbf{S}^{\top}) = \mathbf{E}_R\text{vec}(\mathbf{S}^{\top}).
\end{aligned}
\end{equation}
Using the fact that $\text{tr}(\mathbf{A}^H\mathbf{B})=\text{vec}^H(\mathbf{A})\text{vec}(\mathbf{B})$ and $\text{tr}(\mathbf{ABCD})=\text{vec}^{\top}(\mathbf{D})(\mathbf{A}\otimes\mathbf{C}^{\top})\text{vec}(\mathbf{B}^{\top})$, we have
\begin{equation}
\begin{aligned}
    \text{tr}(\mathbf{Q}_k^H\overline{\mathbf{X}}_R) &= \text{vec}^H(\mathbf{Q}_k)\text{vec}(\overline{\mathbf{X}}_R) \\
    &= \text{vec}^H(\mathbf{Q}_k)\mathbf{E}_R\text{vec}(\mathbf{S}^{\top})
    = -\mathbf{b}_k^H\text{vec}(\mathbf{S}^{\top}),
\end{aligned}
\end{equation}
\begin{equation}
\begin{aligned}
    \text{tr}(\mathbf{T}_k\overline{\mathbf{X}_R}\overline{\mathbf{X}}_R^H) &= \text{tr}(\mathbf{I}_L\overline{\mathbf{X}}_R^H\mathbf{T}_k\overline{\mathbf{X}}_R) \\
    &= \text{vec}^{\top}(\overline{\mathbf{X}}_R)(\mathbf{I}_L\otimes\mathbf{T}_k^*)\text{vec}(\overline{\mathbf{X}}_R^*) \\
    &= \text{vec}(\mathbf{S}^{\top})^{\top}\mathbf{E}_R^{\top}(\mathbf{I}_L\otimes\mathbf{T}_k^*)\mathbf{E}_R^*\text{vec}(\mathbf{S}^{\top})^* \\
    &= \text{vec}(\mathbf{S}^{\top})^H\mathbf{E}_R^H(\mathbf{I}_L\otimes\mathbf{T}_k)\mathbf{E}_R\text{vec}(\mathbf{S}^{\top}) \\
    &= \text{vec}(\mathbf{S}^{\top})^H\mathbf{B}_k\text{vec}(\mathbf{S}^{\top})/\sigma_H^2.
\end{aligned}
\end{equation}
Thus we have
\begin{equation}
\begin{aligned}
    &2\Re\left\{\text{tr}(\mathbf{Q}_k^H\overline{\mathbf{X}}_R)\right\} - \text{tr}(\mathbf{T}_k(\sigma_H^2\overline{\mathbf{X}}_R\overline{\mathbf{X}}_R^H+\sigma_R^2\mathbf{I})) \\
    &= \text{tr}(\sigma_R^2\mathbf{T}_k) + 2\Re\left\{\text{tr}(\mathbf{Q}_k^H\overline{\mathbf{X}}_R)\right\} - \sigma_H^2\text{tr}(\mathbf{T}_k\overline{\mathbf{X}}_R\overline{\mathbf{X}}_R^H) \\
    &= c_k - 2\Re\left\{\text{vec}(\mathbf{S}^{\top})^H\mathbf{b}_k\right\}-\text{vec}(\mathbf{S}^{\top})^H\mathbf{B}_k\text{vec}(\mathbf{S}^{\top}).
\end{aligned}
\end{equation}

\section*{Appendix F: Proof of Proposition 6}
\label{app:prop6}
The proof of the convergence of the BPS algorithm to the optimal solution is presented below. Consider the following convex problem:
\begin{equation}
\label{opt:prov_origin}
\begin{aligned}
    &\underset{\mathbf{x}}{\min} \;
    f_o(\mathbf{x}) \\
    &s.t.\; \mathbf{x}\in\mathcal{S},\; f_p(\mathbf{x})\leq\mathcal{E}.
\end{aligned}
\end{equation}
where $f_o$ and $f_p$ are convex functions and $\mathcal{S}$ is a convex region.
Also consider the penalty problem, which eliminates the energy constraint
\begin{equation}
\label{opt:prov_penalty}
\begin{aligned}
    \mathcal{P}(\rho):\; &\underset{\mathbf{x}}{\min} \;
    f_o(\mathbf{x})+\rho f_p(\mathbf{x}) \\
    &s.t.\; \mathbf{x}\in\mathcal{S}.
\end{aligned}
\end{equation}
Initially, we will prove that as $\rho$ increases, the optimal solution of $\mathcal{P}(\rho)$, $\mathbf{x}^{\star}$, exhibits a larger $f_o(\mathbf{x}^{\star})$ and a smaller $f_p(\mathbf{x}^{\star})$. Subsequently, we will establish that the optimal solution of problem (\ref{opt:prov_origin}) is also the optimal solution of problem $\mathcal{P}(\rho)$ for a certain $\rho$.
\begin{align}
    f_o(\mathbf{x}_1^{\star})+\rho_2f_p(\mathbf{x}_1^{\star})\geq f_o(\mathbf{x}_2^{\star})+\rho_2f_p(\mathbf{x}_2^{\star}), \label{eq:prov1}\\
    f_o(\mathbf{x}_1^{\star})+\rho_1f_p(\mathbf{x}_1^{\star})\leq f_o(\mathbf{x}_2^{\star})+\rho_1f_p(\mathbf{x}_2^{\star}). \label{eq:prov2}
\end{align}
Then we rearrange the inequality (\ref{eq:prov2}) as
\begin{equation}
\label{eq:prov3}
\begin{aligned}
    f_o(\mathbf{x}_1^{\star})+\rho_2f_p(\mathbf{x}_1^{\star})+(\rho_1-\rho_2)f_p(\mathbf{x}_1^{\star}) \\ \leq f_o(\mathbf{x}_2^{\star})+\rho_2f_p(\mathbf{x}_2^{\star})+(\rho_1-\rho_2)f_p(\mathbf{x}_2^{\star}).
\end{aligned}
\end{equation}
According to inequality (\ref{eq:prov1}) and (\ref{eq:prov3}) we have
\begin{equation}
\begin{aligned}
    0\leq\left(f_o(\mathbf{x}_1^{\star})+\rho_2f_p(\mathbf{x}_1^{\star})\right) - \left(f_o(\mathbf{x}_2^{\star})+\rho_2f_p(\mathbf{x}_2^{\star})\right)
    \\ \leq
    (\rho_1-\rho_2)(f_p(\mathbf{x}_2^{\star})-f_p(\mathbf{x}_1^{\star})),
\end{aligned}
\end{equation}
which yields
\begin{equation}
\label{eq:prov4}
    f_p(\mathbf{x}_1^{\star})\leq f_p(\mathbf{x}_2^{\star}).
\end{equation}
Then combining inequality (\ref{eq:prov1}) and (\ref{eq:prov4}) we have
\begin{equation}
\label{eq:prov5}
    f_o(\mathbf{x}_1^{\star})\geq f_o(\mathbf{x}_2^{\star})
\end{equation}
Inequality (\ref{eq:prov4}) and (\ref{eq:prov5}) reveal that an increase in the penalty factor results in an increment of the objective function and a reduction in the penalty function. The objective of the BPS algorithm is to identify the minimal penalty factor such that the penalty function $f_p$ does not surpass $\mathcal{E}$.

We now prove that the optimal solution $\mathbf{x}^{\star}$ of problem (\ref{opt:prov_origin}) corresponds to the optimal solution of problem $\mathcal{P}(\rho)$ for a particular $\rho$. Assuming $\mathbf{x}^{\star}$ is not constrained by $f_p(\mathbf{x})\leq\mathcal{E}$, it is evident that $\mathbf{x}^{\star}$ represents the optimal solution of $\mathcal{P}(0)$. If it is constrained, in accordance with the KKT complementary slackness condition, we ascertain that for certain values of $\mu$ and $\bm{\nu}$:
\begin{equation}
    \frac{\partial f_o}{\partial\mathbf{x}}(\mathbf{x}^{\star})+\mu\frac{\partial f_p}{\partial\mathbf{x}}(\mathbf{x}^{\star})+\frac{\partial{\mathcal{L}_{\mathcal{S}}(\bm{\nu})}}{\partial\mathbf{x}}(\mathbf{x}^{\star})=0,
\end{equation}
where $\mathcal{L}_{\mathcal{S}}$ and $\bm{\nu}$ denote the Lagrange augmentation function and dual variables for the constraint $\mathbf{x}\in\mathcal{S}$. Consequently, we deduce that this condition also represents the KKT complementary slackness condition for problem $\mathcal{P}(\mu)$. Therefore, $\mathbf{x}^{\star}$ serves as the optimal solution of $\mathcal{P}(\mu)$.
\bibliographystyle{IEEEtran}
\bibliography{reference}

\begin{thebibliography}{10}
\providecommand{\url}[1]{#1}
\csname url@samestyle\endcsname
\providecommand{\newblock}{\relax}
\providecommand{\bibinfo}[2]{#2}
\providecommand{\BIBentrySTDinterwordspacing}{\spaceskip=0pt\relax}
\providecommand{\BIBentryALTinterwordstretchfactor}{4}
\providecommand{\BIBentryALTinterwordspacing}{\spaceskip=\fontdimen2\font plus
\BIBentryALTinterwordstretchfactor\fontdimen3\font minus
  \fontdimen4\font\relax}
\providecommand{\BIBforeignlanguage}[2]{{%
\expandafter\ifx\csname l@#1\endcsname\relax
\typeout{** WARNING: IEEEtran.bst: No hyphenation pattern has been}%
\typeout{** loaded for the language `#1'. Using the pattern for}%
\typeout{** the default language instead.}%
\else
\language=\csname l@#1\endcsname
\fi
#2}}
\providecommand{\BIBdecl}{\relax}
\BIBdecl

\bibitem{pin2021integrated}
D.~K. Pin~Tan, J.~He, Y.~Li, A.~Bayesteh, Y.~Chen, P.~Zhu, and W.~Tong,
  ``Integrated sensing and communication in 6{G}: Motivations, use cases,
  requirements, challenges and future directions,'' in \emph{2021 1st IEEE
  International Online Symposium on Joint Communications \& Sensing (JC\&S)},
  2021, pp. 1--6.

\bibitem{zheng2019radar}
L.~Zheng, M.~Lops, Y.~C. Eldar, and X.~Wang, ``Radar and communication
  coexistence: An overview: A review of recent methods,'' \emph{IEEE Signal
  Processing Magazine}, vol.~36, no.~5, pp. 85--99, 2019.

\bibitem{liu2022integrated}
F.~Liu, Y.~Cui, C.~Masouros, J.~Xu, T.~X. Han, Y.~C. Eldar, and S.~Buzzi,
  ``Integrated sensing and communications: Toward dual-functional wireless
  networks for 6{G} and beyond,'' \emph{IEEE Journal on Selected Areas in
  Communications}, vol.~40, no.~6, pp. 1728--1767, 2022.

\bibitem{feng2020china}
Z.~Feng, Z.~Fang, Z.~Wei, X.~Chen, Z.~Quan, and D.~Ji, ``Joint radar and
  communication: A survey,'' \emph{China Communications}, vol.~17, no.~1, pp.
  1--27, 2020.

\bibitem{li2007mimo}
J.~Li and P.~Stoica, ``{MIMO} radar with colocated antennas,'' \emph{IEEE
  Signal Processing Magazine}, vol.~24, no.~5, pp. 106--114, 2007.

\bibitem{mccormick2017simultaneous}
P.~M. McCormick, S.~D. Blunt, and J.~G. Metcalf, ``Simultaneous radar and
  communications emissions from a common aperture, part i: Theory,'' in
  \emph{2017 IEEE Radar Conference (RadarConf)}.\hskip 1em plus 0.5em minus
  0.4em\relax IEEE, 2017, pp. 1685--1690.

\bibitem{qian2018joint}
J.~Qian, M.~Lops, L.~Zheng, X.~Wang, and Z.~He, ``Joint system design for
  coexistence of {MIMO} radar and {MIMO} communication,'' \emph{IEEE
  Transactions on Signal Processing}, vol.~66, no.~13, pp. 3504--3519, 2018.

\bibitem{tang2020waveform}
B.~Tang, H.~Wang, L.~Qin, and L.~Li, ``Waveform design for dual-function {MIMO}
  radar-communication systems,'' in \emph{2020 IEEE 11th Sensor Array and
  Multichannel Signal Processing Workshop (SAM)}.\hskip 1em plus 0.5em minus
  0.4em\relax IEEE, 2020, pp. 1--5.

\bibitem{kumari2019adaptive}
P.~Kumari, S.~A. Vorobyov, and R.~W. Heath, ``Adaptive virtual waveform design
  for millimeter-wave joint communication--radar,'' \emph{IEEE Transactions on
  Signal Processing}, vol.~68, pp. 715--730, 2019.

\bibitem{liu2020range}
F.~Liu, C.~Masouros, T.~Ratnarajah, and A.~Petropulu, ``On range sidelobe
  reduction for dual-functional radar-communication waveforms,'' \emph{IEEE
  Wireless Communications Letters}, vol.~9, no.~9, pp. 1572--1576, 2020.

\bibitem{liu2018toward}
F.~Liu, L.~Zhou, C.~Masouros, A.~Li, W.~Luo, and A.~Petropulu, ``Toward
  dual-functional radar-communication systems: Optimal waveform design,''
  \emph{IEEE Transactions on Signal Processing}, vol.~66, no.~16, pp.
  4264--4279, 2018.

\bibitem{liu2018mu}
F.~Liu, C.~Masouros, A.~Li, H.~Sun, and L.~Hanzo, ``{MU}-{MIMO} communications
  with {MIMO} radar: From co-existence to joint transmission,'' \emph{IEEE
  Transactions on Wireless Communications}, vol.~17, no.~4, pp. 2755--2770,
  2018.

\bibitem{cheng2020hybrid}
Z.~Cheng, B.~Liao, and Z.~He, ``Hybrid transceiver design for dual-functional
  radar-communication system,'' in \emph{2020 IEEE 11th Sensor Array and
  Multichannel Signal Processing Workshop (SAM)}.\hskip 1em plus 0.5em minus
  0.4em\relax IEEE, 2020, pp. 1--5.

\bibitem{yuan2020bayesian}
W.~Yuan, F.~Liu, C.~Masouros, J.~Yuan, D.~W.~K. Ng, and
  N.~Gonz{\'a}lez-Prelcic, ``Bayesian predictive beamforming for vehicular
  networks: A low-overhead joint radar-communication approach,'' \emph{IEEE
  Transactions on Wireless Communications}, vol.~20, no.~3, pp. 1442--1456,
  2020.

\bibitem{xu2020multi}
C.~Xu, B.~Clerckx, and J.~Zhang, ``Multi-antenna joint radar and
  communications: Precoder optimization and weighted sum-rate vs probing power
  tradeoff,'' \emph{IEEE Access}, vol.~8, pp. 173\,974--173\,982, 2020.

\bibitem{su2020secure}
N.~Su, F.~Liu, and C.~Masouros, ``Secure radar-communication systems with
  malicious targets: Integrating radar, communications and jamming
  functionalities,'' \emph{IEEE Transactions on Wireless Communications},
  vol.~20, no.~1, pp. 83--95, 2020.

\bibitem{bazzi2023outage}
A.~Bazzi and M.~Chafii, ``On outage-based beamforming design for
  dual-functional radar-communication 6g systems,'' \emph{IEEE Transactions on
  Wireless Communications}, 2023.

\bibitem{anderson2013faster}
J.~B. Anderson, F.~Rusek, and V.~{\"O}wall, ``Faster-than-{N}yquist
  signaling,'' \emph{Proceedings of the IEEE}, vol. 101, no.~8, pp. 1817--1830,
  2013.

\bibitem{liu2020joint1}
X.~Liu, T.~Huang, N.~Shlezinger, Y.~Liu, J.~Zhou, and Y.~C. Eldar, ``Joint
  transmit beamforming for multiuser {MIMO} communications and {MIMO} radar,''
  \emph{IEEE Transactions on Signal Processing}, vol.~68, pp. 3929--3944, 2020.

\bibitem{chen2020composite}
L.~Chen, F.~Liu, J.~Liu, and C.~Masouros, ``Composite signalling for {DFRC}:
  Dedicated probing signal or not?'' \emph{arXiv preprint arXiv:2009.03528},
  2020.

\bibitem{liu2020joint2}
F.~Liu and C.~Masouros, ``Joint beamforming design for extended target
  estimation and multiuser communication,'' in \emph{2020 IEEE Radar Conference
  (RadarConf20)}.\hskip 1em plus 0.5em minus 0.4em\relax IEEE, 2020, pp. 1--6.

\bibitem{spano2018faster}
D.~Spano, M.~Alodeh, S.~Chatzinotas, and B.~Ottersten, ``Faster-than-{N}yquist
  signaling through spatio-temporal symbol-level precoding for the multiuser
  miso downlink channel,'' \emph{IEEE Transactions on Wireless Communications},
  vol.~17, no.~9, pp. 5915--5928, 2018.

\bibitem{masouros2009dynamic}
C.~Masouros and E.~Alsusa, ``Dynamic linear precoding for the exploitation of
  known interference in {MIMO} broadcast systems,'' \emph{IEEE Transactions on
  Wireless Communications}, vol.~8, no.~3, pp. 1396--1404, 2009.

\bibitem{masouros2015exploiting}
C.~Masouros and G.~Zheng, ``Exploiting known interference as green signal power
  for downlink beamforming optimization,'' \emph{IEEE Transactions on Signal
  Processing}, vol.~63, no.~14, pp. 3628--3640, 2015.

\bibitem{alodeh2018symbol}
M.~Alodeh, D.~Spano, A.~Kalantari, C.~G. Tsinos, D.~Christopoulos,
  S.~Chatzinotas, and B.~Ottersten, ``Symbol-level and multicast precoding for
  multiuser multiantenna downlink: A state-of-the-art, classification, and
  challenges,'' \emph{IEEE Communications Surveys \& Tutorials}, vol.~20,
  no.~3, pp. 1733--1757, 2018.

\bibitem{li2020tutorial}
A.~Li, D.~Spano, J.~Krivochiza, S.~Domouchtsidis, C.~G. Tsinos, C.~Masouros,
  S.~Chatzinotas, Y.~Li, B.~Vucetic, and B.~Ottersten, ``A tutorial on
  interference exploitation via symbol-level precoding: Overview,
  state-of-the-art and future directions,'' \emph{IEEE Communications Surveys
  \& Tutorials}, vol.~22, no.~2, pp. 796--839, 2020.

\bibitem{liu2020joint3}
R.~Liu, M.~Li, Q.~Liu, and A.~L. Swindlehurst, ``Joint symbol-level precoding
  and reflecting designs for irs-enhanced mu-miso systems,'' \emph{IEEE
  Transactions on Wireless Communications}, vol.~20, no.~2, pp. 798--811, 2020.

\bibitem{liu2021intelligent}
R.~Liu, M.~Li, Q.~Liu, A.~L. Swindlehurst, and Q.~Wu, ``Intelligent reflecting
  surface based passive information transmission: A symbol-level precoding
  approach,'' \emph{IEEE Transactions on Vehicular Technology}, vol.~70, no.~7,
  pp. 6735--6749, 2021.

\bibitem{garmatyuk2009wideband}
D.~Garmatyuk, J.~Schuerger, K.~Kauffman, and S.~Spalding, ``Wideband ofdm
  system for radar and communications,'' in \emph{2009 IEEE radar
  conference}.\hskip 1em plus 0.5em minus 0.4em\relax IEEE, 2009, pp. 1--6.

\bibitem{immoreev2002ultra}
I.~I. Immoreev and P.~D.~V. Fedotov, ``Ultra wideband radar systems: advantages
  and disadvantages,'' in \emph{2002 IEEE Conference on Ultra Wideband Systems
  and Technologies (IEEE Cat. No. 02EX580)}.\hskip 1em plus 0.5em minus
  0.4em\relax IEEE, 2002, pp. 201--205.

\bibitem{sen2010multiobjective}
S.~Sen, G.~Tang, and A.~Nehorai, ``Multiobjective optimization of {OFDM} radar
  waveform for target detection,'' \emph{IEEE Transactions on Signal
  Processing}, vol.~59, no.~2, pp. 639--652, 2010.

\bibitem{kay1993fundamentals}
S.~M. Kay, \emph{Fundamentals of Statistical Signal Processing: Estimation
  Theory}.\hskip 1em plus 0.5em minus 0.4em\relax Prentice-Hall, Inc., 1993.

\bibitem{tang2021constrained}
B.~Tang, J.~Liu, H.~Wang, and Y.~Hu, ``Constrained radar waveform design for
  range profiling,'' \emph{IEEE Transactions on Signal Processing}, vol.~69,
  pp. 1924--1937, 2021.

\bibitem{vitthaladevuni2001ber}
P.~K. Vitthaladevuni and M.-S. Alouini, ``Ber computation of 4/m-qam
  hierarchical constellations,'' \emph{IEEE Transactions on Broadcasting},
  vol.~47, no.~3, pp. 228--239, 2001.

\end{thebibliography}

\end{document}